\documentclass[aps,floatfix,superscriptaddress,reprint,10pt,pra]{revtex4-1}
\usepackage{bm}
\usepackage{amsmath}
\usepackage{amssymb}
\usepackage{multirow}
\usepackage{empheq}
\usepackage{graphicx}
\usepackage{mathrsfs}
\usepackage{amsfonts}
\usepackage{amsthm}
\usepackage{color}
\usepackage{bigints}
\usepackage{txfonts}
\usepackage{hyperref}
\hypersetup{
     unicode=false,          
     pdftoolbar=true,        
     pdfmenubar=true,        
     pdffitwindow=false,     
     pdfstartview={FitH},    
     pdftitle={My title},    
     pdfauthor={Author},     
     pdfsubject={Subject},   
     pdfcreator={Creator},   
     pdfproducer={Producer}, 
     pdfkeywords={keyword1} {key2} {key3}, 
     pdfnewwindow=true,      
     colorlinks=false,       
     linkcolor=red,          
     citecolor=green,        
     filecolor=magenta,      
     urlcolor=cyan           
}
\makeatletter \tolerance = 10000 \tolerance = 10000
\usepackage{color}

\makeatother
\begin{document}

\title{Dynamic Viscosity of the ABC-stacked Multilayer Graphene in the Collisionless Regime}

\author{Weiwei Chen}
\affiliation{Key Laboratory of Intelligent Manufacturing Quality Big Data Tracing and Analysis of Zhejiang Province, College of Science, China Jiliang University, Hangzhou 310018, China}
\affiliation{School of Science, Westlake University, 18 Shilongshan Road, Hangzhou 310024, China}

\author{Yedi Shen}
\affiliation{Hefei National Laboratory for Physical Sciences at the Microscale, and Synergetic Innovation Center of Quantum Information and Quantum Physics, University of Science and Technology of China, Hefei, Anhui 230026, China}

\author{Tianle Zhan}
\affiliation{Key Laboratory of Intelligent Manufacturing Quality Big Data Tracing and Analysis of Zhejiang Province, College of Science, China Jiliang University, Hangzhou 310018, China}

\author{W. Zhu}
\thanks{zhuwei@westlake.edu.cn}
\affiliation{School of Science, Westlake University, 18 Shilongshan Road, Hangzhou 310024, China}
\affiliation{Institute of Natural Sciences, Westlake Institute for Advanced Study, 18 Shilongshan Road, Hangzhou 310024,  China}

\date{\today}
\begin{abstract}
We explore the dynamic shear viscosity of the undoped ABC-stacked multilayer graphene based on the chiral-$N$ effective Hamiltonian, where the chirality $N$ is equivalent to the layer number. We investigate the dependence of the dynamic shear viscosity on the frequency in the collisionless regime and calculate Coulomb interaction corrections by three leading order Feynman diagrams: self-energy diagram, vertex diagram, and honey diagram. We propose that the dynamic shear viscosity is generated by the relaxation of momentum flux polarization through electron-hole excitations, and that the interaction can amplify this effect. Furthermore, our research indicates that the dynamic shear viscosity exhibits a robust linear positive dependence on $N$. This finding suggests that by making modifications to the number of layers in graphene, it is possible to finely tune the electron viscous effects.
\end{abstract}
\maketitle

\section{Introduction}
\label{sec:introduction}

Research on the hydrodynamic effects on electrons started long ago.  \cite{Steinberg,Gurzhi1963,Gurzhi1965,Gurzhi1968,Muller2009prl}. It was only recently that experimental observations of viscous electron flows were achieved \cite{Polini2020phytoday,Bandurin2016science,Crossno2016science}. These experiments rapidly ignited interest in the in-depth study of electron hydrodynamics, both theoretically \cite{Lucas2018jpcm,Narozhny2017} and experimentally \cite{Moll2016science,Braem2018prb,Varnavides2020,Vool2021natphy,Jaoui2021natcomm,Gupta2021prl,Bandurin2018NatComm}. Graphene stands as the premier material for observing viscous effects in electrons, owing to its high electron mobility, two-dimensional structure, superior quality, and adjustable electronic properties. These features significantly suppress the loss of total electron momentum, which is crucial for the fluid-like behavior of electrons. Moreover, advanced many-body calculations indicate that the hydrodynamics in AB-stacked bilayer graphene may be more pronounced than in its monolayer counterpart \cite{Tan2022SA}. Therefore, multilayer graphene, with its additional controllable parameter—the number of layers—potentially provides a promising pathway for future exploration of electronic fluid dynamics \cite{Mak2010prl,Wagner2020prb}.

In the study of arbitrarily stacked $N$-layer graphene, it has been found that in the low-energy regime, the spectrum distinctively splits into $N_D \leq N$ separate pseudospin doublets. Each doublet exhibits chiral symmetry with a chirality $J_n$, adhering to the sum rule $\sum_{n=1}^{N_D}J_n=N$ \cite{Min2009prl,Min2008prb}. This relationship underscores the intricate electronic structure of multilayer graphene systems. Notably, $N$-layer ABC-stacked graphene is characterized by a single chiral Hamiltonian, in which the chirality directly equals $N$ [shown in Eq.~(\ref{eq:H0})]. Furthermore, the exploration of charge transport in this system has intriguingly uncovered that, at the low-frequency limit, its optical conductivity tends to be $N$ times that of monolayer graphene \cite{Min2009prl}. Considering the strong linkage between momentum and charge transport, this finding inspires us to investigate how the chirality $N$ affects the viscosity, which characterizes the rate of momentum transport.


Another inevitable issue when discussing electron viscosity is electron-electron interaction. Scattering processes such as electron-impurity, electron-phonon, and electron-boundary can lead to momentum relaxation, whereas the electron-electron interaction preserves momentum conservation, enabling electrons to form collective motion. Thus, in practice, viscous electron flow is expected to be observed in the interaction-dominated regime \cite{Bandurin2016science,Kumar2017natphys,Polini2020phytoday,Muller2009prl}. However, the electron flow experimentally observed in the graphene is detected non-hydrodynamic under condition interaction-dominated but at low temperature, and only exhibits viscous signatures at relatively high temperature \cite{Bandurin2018NatComm}. A plausible interpretation of this phenomenon suggests that the thermal excitation of electrons is critical for the viscous effect. Theoretical calculation concerning disordered Dirac electrons also yield similar prediction that the shear viscosity is suppressed by the impurity scattering but enhanced by the dynamic frequency, which leads to the electron-hole excitation \cite{Chen2022prb}.




In this work, we investigate the dynamic shear viscosity of the ABC-stacked multilayer graphene with neutral charge, where the dynamic frequency will induce electron exicitations analogous to the effects of thermal excitation. We calculate the dynamic shear viscosity of the pure system using the bubble diagram of the Kubo formula for viscosity. The effect of the Coulomb interaction is analyzed by calculating three leading order Feynman diagrams: self-energy diagram, vertex diagram, and honey diagram. This perturbation calculation is valid in the collisionless regime $\Omega\tau_{ee}\gg1$ \cite{Link2018}, where $\Omega$ denotes the dynamic frequency and $\tau_{ee}$ denotes the interaction relaxation time. We find that the dynamic shear viscosity of the pure system satisfies $\eta(\Omega)\propto N\Omega^{2/N}$, exhibiting a power-law dependence on the dynamic frequency with an exponent determined by the chirality $N$. In the high-layer limit, the viscosity tends to be proportional to the number of layers. 
Moreover, the results of interaction correction are positive for any number of layers. 

The paper is organized as follows. In Sec.~\ref{sec:hamilton}, we present the model we considered and the method we used. We review the effective Hamiltonian of the ABC-stacked multilayer graphene and the the Coulomb electron-electron interaction with a soft-cutoff regularization in Sec.~\ref{sec:hamilton}, and show the Kubo formula for viscosity based on the stress-stress correlation function in Sec.~\ref{sec:kubo formula}. Then, the pure dynamic shear viscosity obtained from the bubble diagram is shown in Sec.~\ref{sec:free-model}, combining with a discussion of two different mechanism relaxing the momenum flux: collision-dominant and collisionless. In Sec.~\ref{sec:leading-order-correction}, the correction of dynamic shear viscosity from the Coulomb interaction is calculated by the leading order of the Feynman diagrams: diagram of self-energy correction, diagram of vertex correction, and honey diagram. Finally, we make a conclusion and offer some perspectives in Sec.~\ref{sec:summary and discussion}.

	
	

\section{Model and Methods}
\label{sec:methods}

\subsection{Hamiltonian}
\label{sec:hamilton}

\begin{figure}
	\centering
	\includegraphics[width=0.9\linewidth]{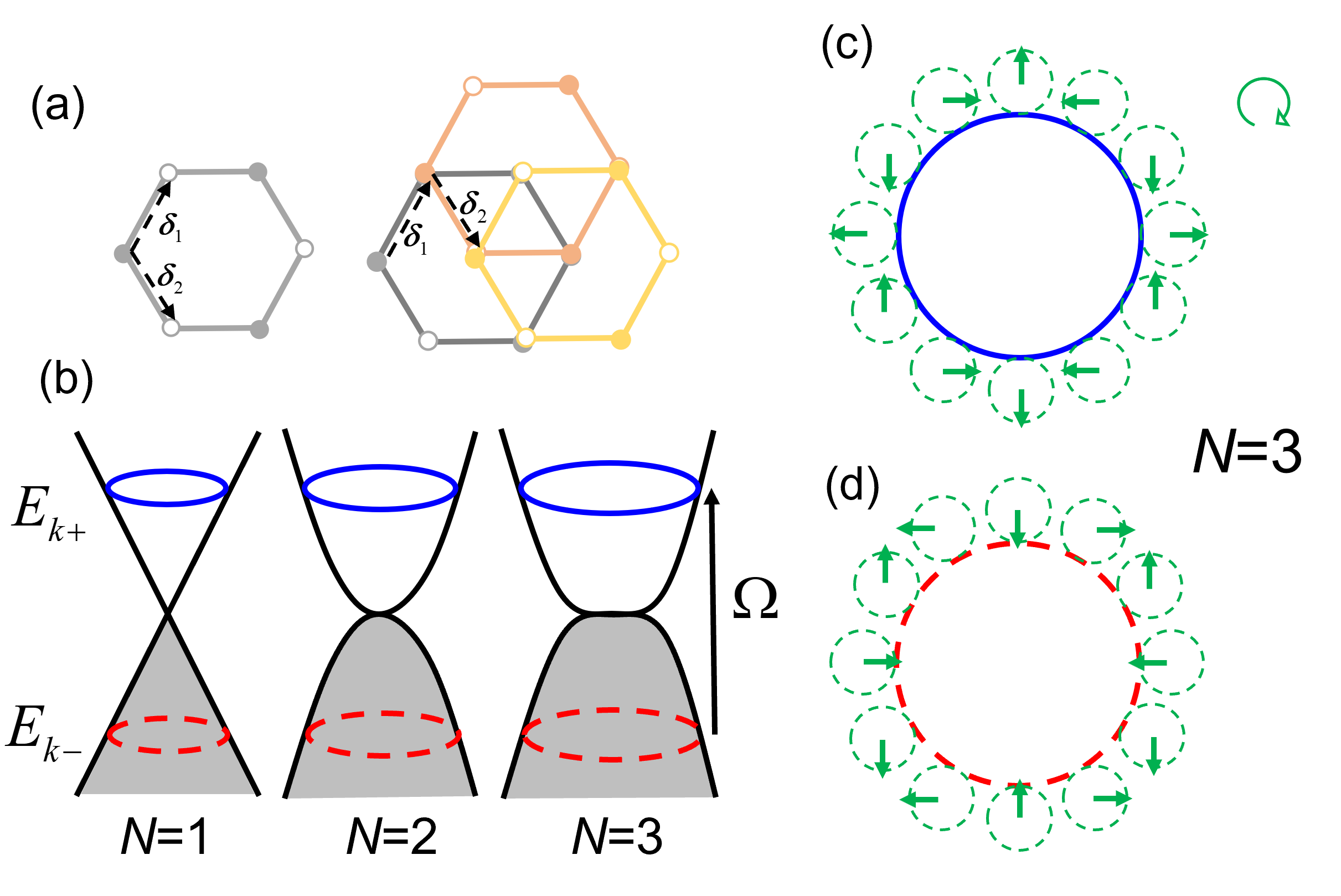}
	\caption{(Color online) 
		(a) Schematic illustration of the periodic ABC stacking arrangement. Filled and open circles represent atoms in two different sublattices. $\bm{\delta}_{1/2}$ denotes vector between nearest atoms. The relative displacement between the first layer and second layer is $\bm{\delta}_1$ while that between the second layer and third layer is $\bm{\delta}_2$. (b) Band structure of the low-energy states of the ABC stacked multilayer graphene. Fermi energy is located at the neutral point. Frequency $\Omega$ induces transition between electron and hole states. 
		(c-d) The pseudospin orientation (green arrows) of electron states ($E_{\bm{k}+}$, blue solid circle) and hole states ($E_{\bm{k}-}$, red dashed circle) in the case of $N=3$.
	}
	\label{fig:sketch-of-abc-stacking}
\end{figure}

In this work, we consider the multilayer graphene with a periodic ABC stacking arrangement depicted in Fig. \ref{fig:sketch-of-abc-stacking}(a). The surface states of an $N$-layer ABC stacked graphene at the low energy can be effectively described by a $N$-chiral fermion with the form \cite{Min2011prb,Min2012prb,Gelderen2013prb}
\begin{equation}\label{eq:H0}
	H_0=\zeta_N\left(\begin{array}{cc}
		0&k_+^N\\k_-^N&0
	\end{array}\right)=\zeta_N(k_+^N\sigma_++k_-^N\sigma_-).
\end{equation}
Here $k_{\pm}=k_x\pm ik_y$, $\bm{k}=(k_x,k_y)$ is a two-component particle momentum. $\sigma_{\pm}=(\sigma_x\pm i\sigma_y)/2$, where $\sigma_{x/y}$ is the Pauli matrix. $\zeta_N\equiv v_0^N/t_{\perp}^{N-1}$. $v_0$ is the effective in-plane Fermi velocity related to the nearest-neighbor intralayer hopping $t_{\parallel}\approx3.09\ eV$ by $\hbar v_0=\sqrt{3}at_{\parallel}/2$, where $a=0.246\ nm$ is the lattice constant of monolayer graphene. $t_{\perp}\approx0.39\ eV$ is the nearest-neighbor interlayer hopping. 

The eigenfunction and eigenenergy of the Hamiltonian Eq.~(\ref{eq:H0}) are given by
\begin{equation}
	\psi_{\bm{k}s}=\frac{1}{\sqrt{2}}\left(\begin{array}{c}
		1\\se^{-iN\theta_{\bm{k}}}
	\end{array}\right);
	\ \ \ \ 
	E_{\bm{k}s}=s\zeta_Nk^N
\end{equation}
where $s=\pm 1$ is the index of subband. $s=-1$ corresponds to the hole states and $s=1$ corresponds to the electron states. According to the eigenfunctions, it is obvious that the number of layers $N$ also stands for the chirality of the effective Hamiltonian, which describes the pseudospin orientation varying with momentum orientation. For example in the case of $N=3$, as shown in Fig.~\ref{fig:sketch-of-abc-stacking} (c-d), while the momentum orientation changes by $2\pi$, the pseudospin orientation changes $2\pi N$. Moreover, for different subbands $s=\pm1$, the direction of the pseudospin orientation is opposite. 

We consider the Coulomb interaction with a soft-cutoff regularization as \cite{Link2018,Mishchenko2008epl}
\begin{equation}
	V(\bm{r})=\frac{\alpha_0 r_0^{-\delta}}{r^{1-\delta}}
\end{equation}
Here, $r_0$ is a length scale introduced to preserve the units of the system at finite $\delta$, and one should take the limit $\delta\to0$ at the end. $\alpha_0=e^2/\epsilon$ characterizes the strength of the interaction, where $\epsilon$ is dielectric constant. The Fourier transformation of this regularized Coulomb interaction is
\begin{equation}
	V(\bm{q})=\frac{2\pi\alpha_{\delta}}{q^{1+\delta}}
\end{equation}
where $\alpha_{\delta}=\alpha_0r_0^{-\delta}2^{\delta}\frac{\Gamma[(1+\delta)/2]}{\Gamma[(1-\delta)/2]}$. 

\subsection{Kubo formula for viscosity}
\label{sec:kubo formula}
The shear viscosity, which relates the viscous stress to the strain rate in the same direction, can be calculated by \cite{Chen2022prb,Link2018}
\begin{equation}\label{eq:eta-correlation}
	\eta(\Omega)=\frac{{\rm Im}C_{xyxy}(\Omega)}{\Omega}
\end{equation}
with the stress-stress correlation function
\begin{equation}\label{eq:correlation-TT-1}
	C_{xyxy}(\Omega)=i\int_0^tdt\langle[\hat{T}_{xy}(t),\hat{T}_{xy}(0)]\rangle_0e^{i\Omega^+t}
\end{equation}
where $\langle\cdots\rangle_0$ means the thermodynamic average over the eigenstates of pure Hamiltonian $H_0$.
$\hat{T}_{ij}$ is the operator of stress tensor along $\hat{i}$ acting on the surface element with normal $\hat{j}$, which can derived by the strain transformation of the system and expressed as \cite{Chen2022prb}
\begin{equation}
	\hat{T}_{ij}=\frac{i}{\hbar}[\hat{\mathcal{J}}_{ij},\hat{H}_0+\hat{V}]
\end{equation}
where $\hat{\mathcal{J}}_{ij}=-\frac{1}{2}(x_ik_j+x_jk_i)$ is the strain transformation generator. For the pure system $H_0$, the pure stress tensor operator $\hat{T}_{xy}^{(0)}$ is single-particle and given by
\begin{equation}
	\hat{T}_{xy}^{(0)}=\sum_{\bm{k}}T^{(0)}_{xy}(\bm{k})\psi^{\dagger}_{\bm{k}}\psi_{\bm{k}}
\end{equation}
with
\begin{equation}\label{eq:T0}
	T_{xy}^{(0)}(\bm{k})=i\frac{N\zeta_N}{2}(k_-k_+^{N-1}\sigma_+-k_+k_-^{N-1}\sigma_-)
\end{equation}
However, the interaction corrects stress tensor $\hat{T}_{xy}^{\text{int}}$ is given by two-body interaction
\begin{equation}
	\hat{T}_{xy}^{\text{int}}=\frac{1}{\mathcal{V}}\sum_{\bm{k}\bm{k}'\bm{q}}T_{xy}^{\text{int}}(\bm{q})\psi^{\dagger}_{\bm{k}+\bm{q}}\psi_{\bm{k}}\psi^{\dagger}_{\bm{k}'-\bm{q}}\psi_{\bm{k}'}
\end{equation}
with
\begin{equation}
	T_{xy}^{\text{int}}(\bm{q})=-\frac{(1-\delta)}{2}\alpha_0r_0^{-\delta}2^{\delta+1}\pi\frac{\Gamma(\frac{3+\delta}{2})}{\Gamma(\frac{3-\delta}{2})}\frac{q_xq_y}{q^{3+\delta}}
\end{equation}
where $\mathcal{V}$ denotes the size of sample. In order to calculate the retarded stress-stress correlation function Eq.~(\ref{eq:eta-correlation}), one can first transform it into Matsubara function by analytical continuation $\Omega^+\to i\Omega$ and $it\to\tau$,
\begin{equation}
	C(i\Omega)=\frac{1}{\mathcal{V}}\int_0^{\beta}d\tau\langle{\rm T}_{\tau}\hat{T}_{xy}(\tau)\hat{T}(0)\rangle_0e^{i\Omega\tau}
\end{equation}
where $\beta=\frac{1}{k_BT}$.

\begin{figure}
	\centering
	\includegraphics[width=1.0\linewidth]{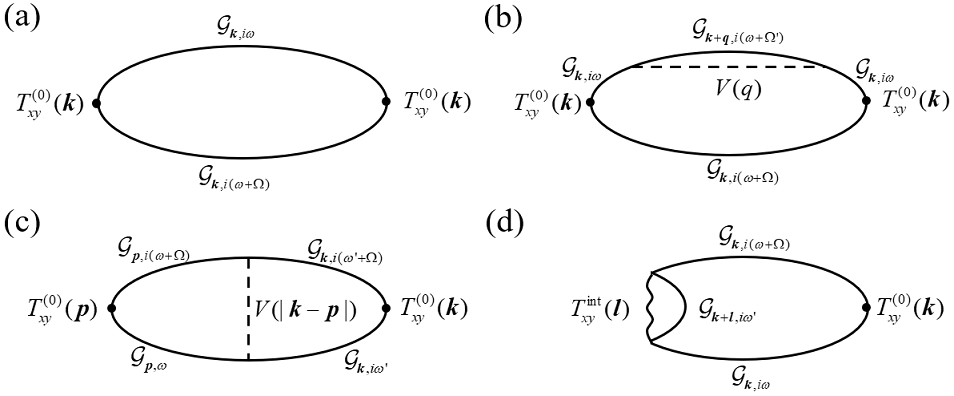}
	\caption{Feynman diagrams for the stress-stress correlation function: (a) bubble diagram; (b) diagram for the self-energy correction; (c) diagram for the vertex correction; (d) honey diagram.}
	\label{fig:sketch-of-feynman-diagrams}
\end{figure}


\begin{figure}
	\centering
	\includegraphics[width=0.9\linewidth]{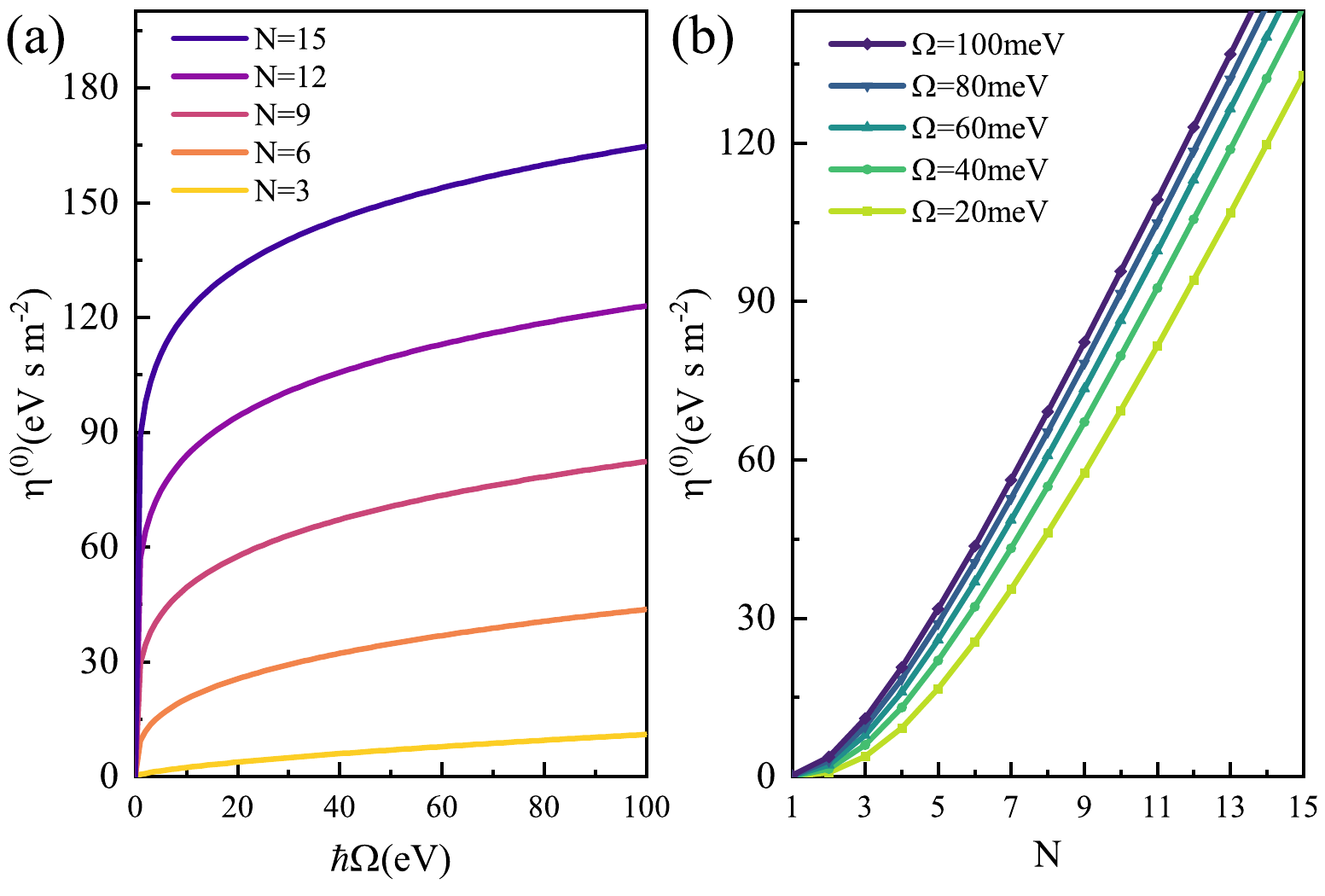}
	\caption{(Color online) The results of pure dynamic shear viscosity $\eta^{(0)}$ obtained by Eq.~(\ref{eq:eta0}) as a function of (a) frequency $\Omega$ and (b) chirality $N$.}
	\label{fig:eta0}
\end{figure}

\section{Dynamic shear viscosity in pure multilayer graphene: the bubble diagram}
\label{sec:free-model}
For the pure ABC-stacked multilayer graphene, the dynamic shear viscosity is derived by the Feynman diagram Fig.~\ref{fig:sketch-of-feynman-diagrams}(a), which is also called the bubble diagram. The expression is given by
\begin{equation}\label{eq:C0}
	C^{(0)}(i\Omega)=-\frac{4}{\beta}\int\frac{d^2\bm{k}}{(2\pi)^2}\sum_{i\omega}{\rm Tr}[T_{xy}^{(0)}(\bm{k})\mathcal{G}_{\bm{k},i\omega+i\Omega}T_{xy}^{(0)}(\bm{k})\mathcal{G}_{\bm{k},i\omega}]
\end{equation}
where 
\begin{equation}
	\mathcal{G}_{\bm{k},i\omega}=\frac{1}{2}\sum_s\frac{1+s\sigma_{\bm{k}}}{E-E_{\bm{k}s}}
	\ \ \ \ \text{with}\ \ \ \
	\sigma_{\bm{k}}=\frac{k_+^N\sigma_++k_-^N\sigma_-}{k^N}
\end{equation}
is the Matsubara Green's function of the pure system. The factor 4 denotes the degeneracy of spin and valley degrees. It is noticed that a closed fermion loop gives a factor $(-1)$. Plugging the pure stress tensor Eq.~(\ref{eq:T0}) into Eq.~(\ref{eq:C0}) and do analytical continuity $i\Omega\to\Omega+i0^+$, the imaginary part of the correlation function is obtained as
\begin{equation}
	{\rm Im}C^{(0)}_{xyxy}(\Omega)=\frac{N}{8\zeta_N^{\frac{2}{N}}}\left(\frac{\Omega}{2}\right)^{\frac{N+2}{N}}
\end{equation}
Plugging this result into Eq.~(\ref{eq:eta-correlation}), the pure dynamic shear viscosity is obtained as
\begin{equation}\label{eq:eta0}
	\eta^{(0)}(\Omega)=\frac{N}{16}\left(\frac{\Omega}{2\zeta_N}\right)^{\frac{2}{N}}=\frac{N}{12a^2}\left(\frac{t_{\perp}}{t_{\parallel}}\right)^2\left(\frac{\Omega}{2t_{\perp}}\right)^{\frac{2}{N}}
\end{equation}

Figure~\ref{fig:eta0} displays the results of pure dynamic shear viscosity $\eta^{(0)}$ as a function of frequency $\Omega$ and chirality $N$. Notably, as the increase of frequency, the viscosity $\eta^{(0)}$ increases sharply at low frequencies and then tends towards saturation at high frequencies. This trend becomes more pronounced in the cases of larger $N$. Comparing with the analyic result Eq.~(\ref{eq:eta0}), it is attributed the power-law exponent in the function between $\eta^{(0)}$ and $\Omega$ becomeing less than 1. Another significant feasure is that the viscosity $\eta^{(0)}$ exhibits a nearly proportional relationship with $N$, which is obviously shown in Fig.~\ref{fig:eta0}(b). Corresponding to the analyic result Eq.~(\ref{eq:eta0}), this reflects that the viscosity approaches $\eta^{(0)}\to\frac{N}{12a^2}\left(\frac{t_{\perp}}{t_{\parallel}}\right)^2$ at large $N$ limit.




In order to figure out the how a finite dynamic viscosity can come about in a pure system, we also derived the Kubo formula of viscosity in the form of retarded Green's function (details shown in Appendix.~\ref{appendix:Kubo-retarded}).
\begin{equation}\begin{aligned}\label{eq:eta-retarded}
		&\eta_s^{(0)}(\Omega)=\frac{N^2\zeta_N^2}{2\pi}\int_{-\infty}^{\infty}d\omega\frac{f(\omega)-f(\omega+\Omega)}{\Omega}\int\frac{d^2\bm{k}}{(2\pi)^2}k^{2N}\\
		&\left[{\rm Im}G^R_{\bm{k},+}(\omega+\Omega){\rm Im}G^R_{\bm{k},+}(\omega)+{\rm Im}G^R_{\bm{k},-}(\omega+\Omega){\rm Im}G^R_{\bm{k},-}(\omega)\right.\\
		&\left.+{\rm Im}G^R_{\bm{k},+}(\omega+\Omega){\rm Im}G^R_{\bm{k},-}(\omega)+{\rm Im}G^R_{\bm{k},-}(\omega+\Omega){\rm Im}G^R_{\bm{k},+}(\omega)\right]
\end{aligned}\end{equation}
where $f(\omega)=[\exp(\beta\omega)+1]^{-1}$ is the Fermi-Dirac distribution function, $G^R_{\bm{k}\pm}(\omega)=(\omega-E_{\bm{k}\pm}+i\Gamma)^{-1}$ is the retarded Green's function in eigen basis, and $\Gamma$ denotes the scattering rate. 

From the formula in the form of retarded Green's function, we can distinguish two different mechanisms leading to viscosity in the collision-dominant and collisionless regimes. In the collision-dominant condition where $\Gamma\ne0$ and $\Omega\to0$, the viscosity mainly comes from the terms ${\rm Im}G^R_{\bm{k},+}(\omega){\rm Im}G^R_{\bm{k},+}(\omega)$ and ${\rm Im}G^R_{\bm{k},-}(\omega){\rm Im}G^R_{\bm{k},-}(\omega)$, which will lead to $\eta\propto\frac{1}{\Gamma}\propto l$. Here $l$ denotes the mean free path. Since ${\rm Im}G^R_{\bm{k},\pm}(\omega)$ denotes the spectral function which reflects the probability of a quasiparticle, these contributions imply that viscosity originates from the momentum transfer through quasiparticle collisions, which is consistent with the classical interpretation of viscosity as "diffusion of momentum" \cite{LaudauVol10}. 

In the collisionless condition where $\Gamma\to0^+$ and $\Omega\ne0$, the viscosity mainly comes from the terms ${\rm Im}G^R_{\bm{k},+}(\omega+\Omega){\rm Im}G^R_{\bm{k},-}(\omega)$ and ${\rm Im}G^R_{\bm{k},-}(\omega+\Omega){\rm Im}G^R_{\bm{k},+}(\omega)$, which implies that frequency $\Omega$ induces the transition between electron and hole states. At the zero temperature and collisionless regime $\Gamma\to0^+$, the Eq.~(\ref{eq:eta0}) can be derived in another form, $\eta\propto\bar{T}^2\rho$, where $\bar{T}$ denotes momentum flux and $\rho$ denotes the density of state (details shown in Appendix.~\ref{appendix:Kubo-retarded}). This result roughly demonstrates the momentum flux polarization relaxed by the electron-hole excitations.


\section{Leading Order Coulomb Interaction Correction}
\label{sec:leading-order-correction}

In the following, we analyze the effect of the Coulomb interaction by calculating the leading order of the Feynman diagrams: diagram of self-energy correction, diagram of vertex correction, and honey diagram, which are shown in the Fig.~\ref{fig:sketch-of-feynman-diagrams}(b), (c), and (d), respectively. Among them, the honey diagram comes from the correction function including interaction-corrected stress tensor, i.e. $\langle{\rm T}_{\tau}T_{xy}^{\text{int}}(\tau)T_{xy}^{(0)}(0)\rangle_0$. Due to the divergence in the limit $\delta\to0$, the calculation should be performed separating into two cases: $N=1$ and $N\geq2$. Since the case for $N=1$ aligns with the findings in Ref~\cite{Link2018}, where the dynamic shear viscosity of monolayer graphene is discussed, we will focus on scenarios where $N\geq2$.

\subsection{Self-energy correction}
\label{sec:self-energy correction}

At first, we calculate the correlation function $C_{xyxy}^{\text{self}}(i\Omega)$ which represents the self-energy correction depicted in the Fig.~\ref{fig:sketch-of-feynman-diagrams}(b), the expression of which is given by
\begin{equation}\begin{aligned}\label{eq:correltaion-self-1}
	C_{xyxy}^{\text{self}}(i\Omega)=&-8\int\frac{d^2\bm{k}}{(2\pi)^2}\frac{d\omega}{2\pi}\\
	&{\rm Tr}[\mathcal{G}_{\bm{k}i(\omega+\Omega)}T_{xy}^{(0)}(\bm{k})\mathcal{G}_{\bm{k},i\omega}\Sigma(\bm{k})\mathcal{G}_{\bm{k},i\omega}T_{xy}^{(0)}(\bm{k})]
\end{aligned}\end{equation}
where the factor $8=2\times4$ with $2$ from the divergence of Feynman diagram and $4$ from divergence of spin and valley. The self-energy function $\Sigma(\bm{k})$ is given by  (details shown in Appendix~\ref{appendix:self-energy correlation})
\begin{equation}\label{eq:self}
	\Sigma(\bm{k})=-\frac{1}{\beta\mathcal{V}}\sum_{\bm{q},i\Omega'}V(q)\mathcal{G}_{\bm{k}+\bm{q},i(\omega+\Omega')}=\phi(k)\sigma_{\bm{k}}
\end{equation}
where
\begin{equation}\label{eq:self-phi}
	\phi(k)=\alpha_0r_0^{-\delta}2^{\delta-3}\frac{\Gamma(\frac{N-1+\delta}{2})}{\Gamma(\frac{N+3-\delta}{2})}k^{1-\delta}
\end{equation}
Here, $\Gamma(x)$ is gamma function. Plugging Eq.~(\ref{eq:self}) and Eq.~(\ref{eq:self-phi}) into Eq.~(\ref{eq:correltaion-self-1}), the imaginary part of the self-energy correlated correlation function is obtained as
\begin{equation}\label{eq:correltaion-self-n-2}
	{\rm Im}C_{xyxy}^{\text{self}}(i\Omega)={\rm Im}\left[\alpha_{0}\frac{N(N+3)}{2^{6+\frac{3}{N}}}\frac{\Gamma(\frac{N-1}{2})}{\Gamma(\frac{N+3}{2})}\csc(\frac{3\pi}{2N})\zeta_N^{-\frac{3}{N}}\Omega^{\frac{3}{N}}\right]
\end{equation}
where $d=2+\frac{3}{N}$ and $N\geq2$. 

\subsection{Vertex correction}
\label{sec:vertex-correction}
Next, we move on to the vertex-correction part of the correlation function $C_{xyxy}^{\text{ver}}(i\Omega)$ depicted in Fig.~\ref{fig:sketch-of-feynman-diagrams}(c), which is given by
\begin{equation}\begin{aligned}
	C_{xyxy}^{\text{ver}}(i\Omega)=&4\int\frac{d\omega}{2\pi}\int\frac{d\omega'}{2\pi}\int\frac{d^2\bm{p}}{(2\pi)^2}\int\frac{d^2\bm{k}}{(2\pi)^2}V(|\bm{k}-\bm{p}|)\\
	&{\rm Tr}[\mathcal{G}_{\bm{p},i\omega}T_{xy}^{(0)}(\bm{p})\mathcal{G}_{\bm{p},i(\omega+\Omega)}\mathcal{G}_{\bm{k},i(\omega'+\Omega)}T_{xy}^{(0)}(\bm{k})\mathcal{G}_{\bm{k},i\omega'}]
\end{aligned}\end{equation}
In the case of $N\geq2$, this equation is derived by dividing into three parts (details shown in Appendix~\ref{appendix:vertex}),
\begin{equation}
	{\rm Im}\mathcal{C}_{xyxy}^{\text{ver}}(i\Omega)={\rm Im}[Q^{\text{ver}}_1+Q^{\text{ver}}_2+Q^{\text{ver}}_3]
\end{equation}
where
\begin{equation}\begin{aligned}
		Q_1^{\text{ver}}=&-\frac{N}{2^{4+\frac{3}{N}}\pi}\csc(\frac{3\pi}{2N})f_1(N)\zeta_N^{-\frac{3}{N}}\alpha_{0}\Omega^{\frac{3}{N}}\\
		Q_2^{\text{ver}}=&-\frac{N}{2^{4+\frac{3}{N}}\pi}\csc(\frac{3\pi}{2N})f_2(N)\zeta_N^{-\frac{3}{N}}\alpha_{0}\Omega^{\frac{3}{N}}\\
		Q_3^{\text{ver}}=&-\frac{N}{2^{5+\frac{3}{N}}}\csc(\frac{3\pi}{2N})g_1(N)\zeta_N^{-\frac{3}{N}}\alpha_0\Omega^{\frac{3}{N}}\\
\end{aligned}\end{equation}
with
\begin{equation}\begin{aligned}
	f_1(N)=&\int_0^{\infty} dx\int_0^{\pi}d\theta\frac{x^{N+1}\cos(2\theta)}{(x^2+1-2x\cos\theta)^{\frac{1}{2}}}\frac{1-x^{-3}}{x^{2N}-1}\\
	f_2(N)=&\int_0^{\infty} dx\int_0^{\pi}d\theta\frac{x\cos(2\theta)\cos(N\theta)}{(x^2+1-2x\cos\theta)^{\frac{1}{2}}}\frac{1-x^{2N-3}}{x^{2N}-1}
\end{aligned}\end{equation}
and
\begin{equation}\label{eq:g1}
	g_1(N)=\int_0^{\infty}dz\frac{e^{-z/2}}{z^{2}}[(z-1+N)I_{\frac{N-1}{2}}(\frac{z}{2})+(z-1-N)I_{\frac{N+1}{2}}(\frac{z}{2})]
\end{equation}

\subsection{Honey diagram correction}
\label{sec:honey}
Then, we calculate the correlation function including interaction-corrected stress tensor, i.e. $C_{xyxy}^{\text{honey}}(i\Omega)$, the Feynman diagram of which is called the ``honey diagram" shown in Fig.~\ref{fig:sketch-of-feynman-diagrams}(d). It is given by
\begin{equation}\begin{aligned}
	C_{xyxy}^{\text{honey}}(i\Omega)=&16\int\frac{d^2\bm{k}}{(2\pi)^2}\int\frac{d^2\bm{l}}{(2\pi)^2}\int\frac{d\omega'}{2\pi}\int\frac{d\omega}{2\pi}T_{xy}^{\text{int}}(\bm{l})\\
	&{\rm Tr}\left[\mathcal{G}_{\bm{k}+\bm{l},i\omega'}\mathcal{G}_{\bm{k},i(\omega+\Omega)}T_{xy}^{(0)}(\bm{k})\mathcal{G}_{\bm{k},i\omega}\right]
\end{aligned}\end{equation}
In the case of $N\geq2$, the above equation is obtained as (details shown in Appendix~\ref{appendix:honey-correction})
\begin{equation}
	C_{xyxy}^{\text{honey}}(i\Omega)=-\alpha_0\frac{N}{2^{5+\frac{3}{N}}}\csc(\frac{3\pi}{2N})g_2(N)\zeta_N^{-\frac{3}{N}}\Omega^{\frac{3}{N}}
\end{equation}
with
\begin{equation}\label{eq:g2}
	g_2(N)=\int_0^{\infty}dz\frac{e^{-z/2}}{z^{2}}[(1-N)I_{\frac{N-1}{2}}(\frac{z}{2})+(1+N)I_{\frac{N+1}{2}}(\frac{z}{2})]
\end{equation}
The integrals in the Eq.~(\ref{eq:g1}) and Eq.~(\ref{eq:g2}) can be analytically solved by summing them together. Thus, we obtain
\begin{equation}\begin{aligned}
		Q_3^{\text{ver}}+\mathcal{C}_{xyxy}^{\text{honey}}=-\alpha_0\frac{4N^2}{2^{5+\frac{3}{N}}(N^2-1)}\csc(\frac{3\pi}{2N})\zeta_N^{-\frac{3}{N}}\Omega^{\frac{3}{N}}
\end{aligned}\end{equation}

\begin{figure}
	\centering
	\includegraphics[width=1.0\linewidth]{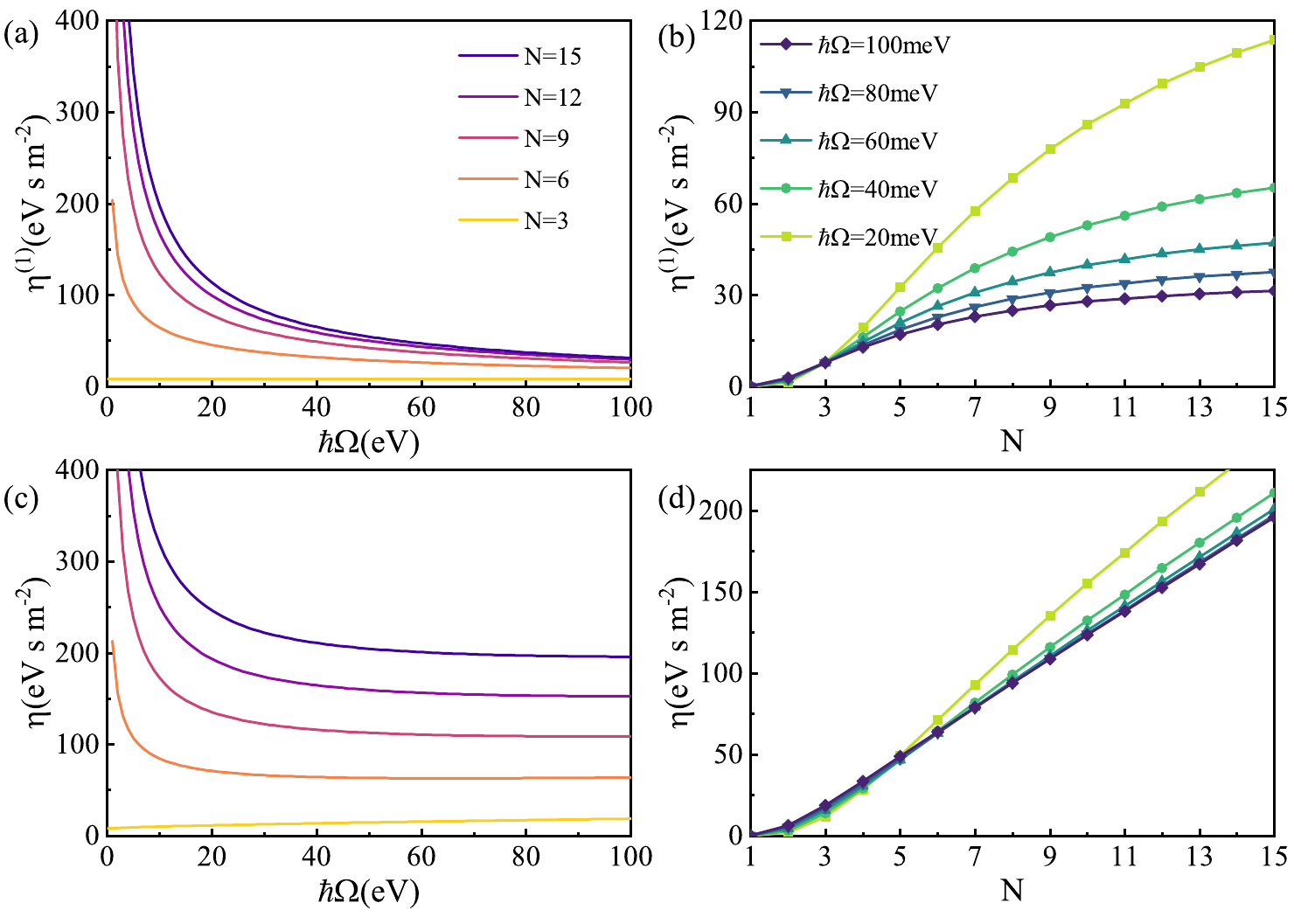}
	\caption{(Color online) The contribution of dynamic shear viscosity from the Coulomb correction $\eta^{(1)}$ as a function of (a) frequency and (b) chirality $N$. The total value of viscosity $\eta$ as a function of (c) frequency and (d) chirality $N$. 
	}
	\label{fig:eta1total}
\end{figure}

\subsection{Total shear viscosity}

Finally, we sum the corrections from three parts and take the limit $\delta\to0$. Then, after taking analytical continuity $i\Omega\to\Omega+i0^+$, which equals $\Omega\to\Omega e^{-i\frac{\pi}{2}+i0^+}$, the imaginary part of leading order correction of the correlation function is obtained as
\begin{equation}\begin{aligned}
		&{\rm Im}C_{xyxy}^{(1)}(\Omega)={\rm Im}[\mathcal{C}_{xyxy}^{\text{self}}(\Omega)+\mathcal{C}_{xyxy}^{\text{ver}}(\Omega)+\mathcal{C}_{xyxy}^{\text{honey}}(\Omega)]\\
		=&\frac{N\alpha_{0}\zeta_N^{-\frac{3}{N}}\Omega^{\frac{3}{N}}}{2^{6+\frac{3}{N}}}\left[-(N+3)\frac{\Gamma(\frac{N-1}{2})}{\Gamma(\frac{N+3}{2})}
		+\frac{4}{\pi}f_1(N)+\frac{4}{\pi}f_2(N)+\frac{8N}{N^2-1}\right]
\end{aligned}\end{equation}

Thus, the leading order interaction-corrected viscosity is given by
\begin{equation}\begin{aligned}
	\eta^{(1)}(\Omega)=&\frac{\tilde{\alpha}N}{96}\left(\frac{t_{\perp}}{t_{\parallel}}\right)^2\left(\frac{\Omega}{2t_{\perp}}\right)^{\frac{3}{N}-1}\frac{\hbar}{a^2}\\
	&\left[-(N+3)\frac{\Gamma(\frac{N-1}{2})}{\Gamma(\frac{N+3}{2})}
	+\frac{4}{\pi}f_1(N)+\frac{4}{\pi}f_2(N)+\frac{8N}{N^2-1}\right]
\end{aligned}\end{equation}
and the total value of dynamic shear viscosity is given by
\begin{equation}
	\eta(\Omega)=\eta^{(0)}(\Omega)+\eta^{(1)}(\Omega)
\end{equation}

The results of the correction part $\eta^{(1)}$ and the total value of dynamic shear viscosity $\eta$ are shown in Fig.~\ref{fig:eta1total}. There are several remarkable features in these results. Firstly, for arbitrary $N$, the interaction correction part $\eta^{(1)}$ is positive. This result is contrary to previous predictions in collision-dominant regime \cite{Muller2009prl,Kiselev2019prb}, which suggested that the viscosity would be suppressed due to the shortening of the mean free path caused by scattering. Combined with the analysis of the pure system, this phenomenon is likely related to the viscous mechanism in th collisionless regime, where Coulomb interaction further enhances electron-hole transitions. Secondly, for a certain $N$, as the increase of frequency $\Omega$, $\eta^{(1)}$ decreases, which is ooposite to the behavior of $\eta^{(0)}$. This opposition makes the total dynamic shear viscosity $\eta$ appear to be independent of $\Omega$, except in the ultralow frequency limit where this perturbation fails. Thirdly, despite the adjustments to the frequency dependence of dynamic shear viscosity due to interaction effects, the relationship with the layer number $N$ continues to exhibit a robust linear dependence.

\section{Conclusion}
\label{sec:summary and discussion}
We have investigated the dynamic shear viscosity in ABC-stacked multilayer graphene based on the chiral-$N$ effective Hamiltonian. Our primary focus is on the frequency and chirality dependencies of the dynamic shear viscosity. We address that the dynamic shear viscosity is generated by the relaxation of momentum flux polarization through electron-hole excitations, and find the interaction can amplify this effect. Moreover, we find that the dynamic shear viscosity exhibits a robust linear positive dependence on the chirality $N$. Since $N$ is equivalent to the layer number of ABC-stacked multilayer graphene, this finding indicates that we can manipulate the electron viscous effect by altering the number of graphene layers.

\begin{acknowledgments}
	We thank Ewelina. M. Hankiewicz for helpful discusstion. W.W.C. and W.Z. are supported by
	National Science Foundation of China (92165102,11974288), and the foundation from Westlake University. This work was supported by ``Pioneer" and ''Leading Goose" R\&D Program of Zhejiang (2022SDXHDX0005), the Key R\&D Program of Zhejiang Province (2021C01002).
\end{acknowledgments}

\begin{widetext}
\begin{appendix}
\section{The pure correlation function}
\subsection{Kubo formula in form of the Matsubara Green's function}
\begin{equation}\begin{aligned}
		\mathcal{C}^0_{xyxy}(i\Omega)=&-\frac{4}{\beta}\int\frac{d^2\bm{k}}{(2\pi)^2}\sum_{i\omega}{\rm Tr}[T_{xy}^{(0)}(\bm{k})\mathcal{G}_{\bm{k},i\omega+i\Omega}T_{xy}^{(0)}(\bm{k})\mathcal{G}_{\bm{k},i\omega}]\\
		=&-\frac{1}{\beta}\int\frac{d^2\bm{k}}{(2\pi)^2}\sum_{i\omega}\sum_{s_1s_2}{\rm Tr}\left[i\frac{N\zeta_N}{2}(k_-k_+^{N-1}\sigma_+-k_+k_-^{N-1}\sigma_-)\frac{1+s_1\sigma_{\bm{k}}}{i(\omega+\Omega)-E_{\bm{k}s_1}}i\frac{N\zeta_N}{2}(k_-k_+^{N-1}\sigma_+-k_+k_-^{N-1}\sigma_-)\frac{1+s_2\sigma_{\bm{k}}}{i\omega-E_{\bm{k}s_2}}\right]\\
		=&\frac{1}{\beta}(\frac{N\zeta_N}{2})^2\int\frac{d^2\bm{k}}{(2\pi)^2}\sum_{i\omega}\sum_{s_1s_2}{\rm Tr}\frac{(k_-k_+^{N-1}\sigma_+-k_+k_-^{N-1}\sigma_-)(1+s_1\sigma_{\bm{k}})(k_-k_+^{N-1}\sigma_+-k_+k_-^{N-1}\sigma_-)(1+s_2\sigma_{\bm{k}})}{[i(\omega+\Omega)-E_{\bm{k}s_1}][i\omega-E_{\bm{k}s_2}]}\\
		=&(\frac{N\zeta_N}{2})^2\int\frac{d^2\bm{k}}{(2\pi)^2}\sum_{s_1s_2}\frac{n_F(E_{\bm{k}s_2})-n_F(E_{\bm{k}s_1})}{i\Omega-E_{\bm{k}s_1}+E_{\bm{k}s_2}}k^4\\
		&{\rm Tr}\left[(k_+^{N-2}\sigma_+-k_-^{N-2}\sigma_-)(1+s_1\frac{k_+^N\sigma_++k_-^N\sigma_-}{k^N})(k_+^{N-2}\sigma_+-k_-^{N-2}\sigma_-)(1+s_2\frac{k_+^N\sigma_++k_-^N\sigma_-}{k^N})\right]\\
		=&(\frac{N\zeta_N}{2})^2\int\frac{d^2\bm{k}}{(2\pi)^2}\sum_{s_1s_2}\frac{n_F(E_{\bm{k}s_2})-n_F(E_{\bm{k}s_1})}{i\Omega-E_{\bm{k}s_1}+E_{\bm{k}s_2}}k^4\\
		&{\rm Tr}\left[(k_+^{N-2}\sigma_+-k_-^{N-2}\sigma_-+s_1\frac{-k_-^{N-2}k_+^N\sigma_-\sigma_++k_+^{N-2}k_-^N\sigma_+\sigma_-}{k^N})(k_+^{N-2}\sigma_+-k_-^{N-2}\sigma_-+s_2\frac{-k_-^{N-2}k_+^N\sigma_-\sigma_++k_+^{N-2}k_-^N\sigma_+\sigma_-}{k^N})\right]\\
		=&(\frac{N\zeta_N}{2})^2\int\frac{d^2\bm{k}}{(2\pi)^2}\sum_{s_1s_2}\frac{n_F(E_{\bm{k}s_2})-n_F(E_{\bm{k}s_1})}{i\Omega-E_{\bm{k}s_1}+E_{\bm{k}s_2}}k^4\\
		&{\rm Tr}\left[-(k_-k_+)^{N-2}(\sigma_+\sigma_-+\sigma_-\sigma_+)+s_1s_2\frac{k_-^{2N-4}k_+^{2N}\sigma_-\sigma_++k_+^{2N-4}k_-^{2N}\sigma_+\sigma_-}{k^{2N}}\right]\\
		=&-2(\frac{N\zeta_N}{2})^2\int\frac{d^2\bm{k}}{(2\pi)^2}\sum_{s_1s_2}\frac{n_F(E_{\bm{k}s_2})-n_F(E_{\bm{k}s_1})}{i\Omega-E_{\bm{k}s_1}+E_{\bm{k}s_2}}k^{2N}\left[1-s_1s_2\frac{k_+^4+k_-^4}{2k^4}\right]\\
		=&-2(\frac{N\zeta_N}{2})^2\int\frac{d^2\bm{k}}{(2\pi)^2}\sum_{s_1s_2}\frac{n_F(E_{\bm{k}s_2})-n_F(E_{\bm{k}s_1})}{i\Omega-E_{\bm{k}s_1}+E_{\bm{k}s_2}}k^{2N}\left[1-s_1s_2\cos(4\theta_{\bm{k}})\right]\\
		=&-(\frac{N\zeta_N}{2})^2\int\frac{kdk}{\pi}\sum_{s_1s_2}\frac{n_F(E_{\bm{k}s_2})-n_F(E_{\bm{k}s_1})}{i\Omega-E_{\bm{k}s_1}+E_{\bm{k}s_2}}k^{2N}\\
\end{aligned}\end{equation}
Then we do analytical continuity $i\Omega\to\Omega+i0^+$ and assume $\Omega\geq0$, so that the imaginary part of the pure correction function is derived as
\begin{equation}\begin{aligned}
		{\rm Im}\mathcal{C}_{xyxy}^0(\Omega)=&-(\frac{N\zeta_N}{2})^2\int\frac{kdk}{\pi}\sum_{s_1s_2}[n_F(E_{\bm{k}s_2})-n_F(E_{\bm{k}s_1})]k^{2N}(-\pi)\delta[\Omega-(s_1-s_2)\zeta_N k^N]\\
		=&(\frac{N\zeta_N}{2})^2\int kdk k^{2N}\delta[\Omega-2\alpha k^N]
		=(\frac{N\zeta_N}{2})^2\int k^{2N+1}dk\frac{\delta(k-\sqrt[N]{\Omega/2\zeta_N})}{2N\zeta_N k^{N-1}}\\
		=&\frac{N\zeta_N}{8}\left(\frac{\Omega}{2\zeta_N}\right)^{\frac{N+2}{N}}
\end{aligned}\end{equation}
so that the pure dynamic shear viscosity is
\begin{equation}\label{app:pure-Mat-eta}
	\eta^{(0)}(\Omega)=\frac{{\rm Im}\mathcal{C}_{xyxy}^0(\Omega)}{\Omega}=\frac{N}{16}\left(\frac{\Omega}{2\zeta_N}\right)^{\frac{2}{N}}=\frac{N}{12a^2}\left(\frac{t_{\perp}}{t_{\parallel}}\right)^2\left(\frac{\Omega}{2t_{\perp}}\right)^{\frac{2}{N}}
\end{equation}

\subsection{Kubo formula in form of the retarded Green's function}\label{appendix:Kubo-retarded}
The Kubo formula of dynamic shear viscosity in the form of retarded and advanced Green's function is given by
\begin{equation}\begin{aligned}\label{evaluation:dynamic-viscosity-1}
		\eta_s(\Omega)=&{\rm Re}\left[\eta_s^{RA}(\Omega)-\eta_s^{RR}(\Omega)\right]
		=\frac{1}{2}\left[\eta_s^{RA}(\Omega)+\eta_s^{AR}(\Omega)-\eta_s^{RR}(\Omega)-\eta_s^{AA}(\Omega)\right]
\end{aligned}\end{equation}
where
\begin{equation}\label{evaluation:dynamic-viscosity-2}
	\eta_s^{LM}(\Omega)=\frac{4}{\mathcal{V}}\int_{-\infty}^{\infty}\frac{d\omega}{2\pi}\frac{f(\omega)-f(\omega+\Omega)}{\Omega}{\rm Tr}\left[G^L(\omega+\Omega)T_{xy}G^M(\omega)T_{xy}\right]
\end{equation}
Combining Eq.~(\ref{evaluation:dynamic-viscosity-1}) with Eq.~(\ref{evaluation:dynamic-viscosity-2}), we can get
\begin{equation}\begin{aligned}\label{evaluation:dynamic-viscosity-3}
		\eta_s^{(0)}(\Omega)=&\frac{4}{\mathcal{V}}\int_{-\infty}^{\infty}\frac{d\omega}{2\pi}\frac{f(\omega)-f(\omega+\Omega)}{\Omega}\frac{-1}{2}{\rm Tr}\left[(G^R_{\omega+\Omega}-G^A_{\omega+\Omega})T_{xy}(G^R_{\omega}-G^A_{\omega})T_{xy}\right]\\
		=&\frac{8}{\mathcal{V}}\int_{-\infty}^{\infty}\frac{d\omega}{2\pi}\frac{f(\omega)-f(\omega+\Omega)}{\Omega}{\rm Tr}\left[{\rm Im}G^R_{\omega+\Omega}T_{xy}{\rm Im}G^R_{\omega}T_{xy}\right]\\
		=&8\int_{-\infty}^{\infty}\frac{d\omega}{2\pi}\frac{f(\omega)-f(\omega+\Omega)}{\Omega}\int\frac{d^2\bm{k}}{(2\pi)^2}\left(\frac{N\zeta_N k^N}{2}\right)^2\\
		&{\rm Tr}
		\left(\begin{array}{cc}
			{\rm Im}G^R_{\bm{k},+}(\omega+\Omega)&\\&{\rm Im}G^R_{\bm{k},-}(\omega+\Omega)
		\end{array}\right)
		\left(\begin{array}{cc}
			\sin2\theta_{\bm{k}} & -i\cos2\theta_{\bm{k}}\\i\cos2\theta_{\bm{k}}&\sin2\theta_{\bm{k}}
		\end{array}\right)
		\left(\begin{array}{cc}
			{\rm Im}G^R_{\bm{k},+}(\omega)&\\&{\rm Im}G^R_{\bm{k},-}(\omega)
		\end{array}\right)
		\left(\begin{array}{cc}
			\sin2\theta_{\bm{k}} & -i\cos2\theta_{\bm{k}}\\i\cos2\theta_{\bm{k}}&\sin2\theta_{\bm{k}}
		\end{array}\right)\\
		=&\frac{N^2\zeta_N^2}{\pi}\int_{-\infty}^{\infty}d\omega\frac{f(\omega)-f(\omega+\Omega)}{\Omega}\int\frac{d^2\bm{k}}{(2\pi)^2}k^{2N}\\
		&{\rm Tr}
		\left(\begin{array}{cc}
			\sin\theta_{\bm{k}}{\rm Im}G^R_{\bm{k},+}(\omega+\Omega)&-i\cos\theta_{\bm{k}}{\rm Im}G^R_{\bm{k},+}(\omega+\Omega)\\i\cos\theta_{\bm{k}}{\rm Im}G^R_{\bm{k},-}(\omega+\Omega)&\sin\theta_{\bm{k}}{\rm Im}G^R_{\bm{k},-}(\omega+\Omega)
		\end{array}\right)
		\left(\begin{array}{cc}
			\sin\theta_{\bm{k}}{\rm Im}G^R_{\bm{k},+}(\omega)&-i\cos\theta_{\bm{k}}{\rm Im}G^R_{\bm{k},+}(\omega)\\i\cos\theta_{\bm{k}}{\rm Im}G^R_{\bm{k},-}(\omega)&\sin\theta_{\bm{k}}{\rm Im}G^R_{\bm{k},-}(\omega)
		\end{array}\right)\\
		=&\frac{N^2\zeta_N^2}{2\pi}\int_{-\infty}^{\infty}d\omega\frac{f(\omega)-f(\omega+\Omega)}{\Omega}\int\frac{d^2\bm{k}}{(2\pi)^2}k^{2N}\\
		&\left[{\rm Im}G^R_{\bm{k},+}{\rm Im}(\omega+\Omega)G^R_{\bm{k},+}(\omega)+{\rm Im}G^R_{\bm{k},-}(\omega+\Omega){\rm Im}G^R_{\bm{k},-}(\omega)+{\rm Im}G^R_{\bm{k},+}(\omega+\Omega){\rm Im}G^R_{\bm{k},-}(\omega)+{\rm Im}G^R_{\bm{k},-}(\omega+\Omega){\rm Im}G^R_{\bm{k},+}(\omega)\right]\\
		\xrightarrow{\Gamma\to0,\Omega\ne0}&\frac{N^2\zeta_N^2}{2\pi}\int_{-\infty}^{\infty}d\omega\frac{f(\omega)-f(\omega+\Omega)}{\Omega}\int\frac{d^2\bm{k}}{(2\pi)^2}k^{2N}\\
		&\pi^2\left[\delta(\omega+\Omega-E_{k,+})\delta(\omega-E_{k,+})+\delta(\omega+\Omega-E_{k,-})\delta(\omega-E_{k,-})+\delta(\omega+\Omega-E_{k,+})\delta(\omega-E_{k,-})+\delta(\omega+\Omega-E_{k,-})\delta(\omega-E_{k,-})\right]\\
		\xrightarrow{\Omega>0}&\frac{N^2\zeta_N^2}{2\pi}\frac{1}{\Omega}\int\frac{d^2\bm{k}}{(2\pi)^2}k^{2N}\pi^2\delta(\Omega+E_{k,-}-E_{k,+})\\
		=&\frac{\pi}{4\Omega}\int\frac{d^2\bm{k}}{(2\pi)^2}(N\zeta_Nk^N)^2\delta(\frac{\Omega}{2}-E_{k,+})\\
		=&\frac{\pi}{2\Omega}\bar{T}^2(\frac{\Omega}{2})\rho(\frac{\Omega}{2})
\end{aligned}\end{equation}
where $f(\omega)=[\exp(\beta\omega)+1]^{-1}$ is the Fermi-Dirac distribution function, $G^R_{\bm{k}\pm}(\omega)=(\omega-E_{\bm{k}\pm}+i\Gamma)^{-1}$ is the retarded Green's function in eigen basis, $\Gamma$ denotes the scattering rate, and  $\bar{T}(\frac{\Omega}{2})=\sqrt{\langle v_x^2k_y^2\rangle}=\frac{1}{2\sqrt{2}}v_{\bm{k}}k\big|_{E=\Omega/2}=\frac{N\Omega}{4\sqrt{2}}$ is the root mean square of momentum flux and $\rho(\frac{\Omega}{2})=\frac{4}{\pi N\Omega}\left(\frac{\Omega}{2\zeta_N}\right)^2$ is the density of state at the  energy level $E_{\bm{k}+}=\frac{\Omega}{2}$. This result is same as that obtained by the Matsubara Green's function, Eq.~(\ref{app:pure-Mat-eta}).

\section{The self-energy correction}\label{appendix:self-energy correlation}
At first, we derive the self-energy function at the first order in pseudospin basis
\begin{equation}\begin{aligned}\label{eq:self-energy-Fork-1}
		\Sigma(\bm{k},ik_n)=&-\frac{1}{\beta\mathcal{V}}\sum_{\bm{q},iq_n}V(\bm{q})\mathcal{G}(\bm{k}+\bm{q},ik_n+iq_n)
		=-\frac{1}{\beta\mathcal{V}}\sum_{\bm{q},iq_n}V(\bm{q})\frac{1}{2}\sum_{s=\pm}\frac{1+s\sigma_{\bm{k}+\bm{q}}}{ik_n+iq_n-E_{\bm{k}+\bm{q},s}}\\
		=&-\frac{1}{\mathcal{V}}\sum_{\bm{q}}V(\bm{q})\frac{1}{2}\sum_{s=\pm}(1+s\sigma_{\bm{k}+\bm{q}})n_F(E_{\bm{k}+\bm{q},s}-\mu)\\
		=&-\frac{1}{\mathcal{V}}\sum_{\bm{p}}V(|\bm{p}-\bm{k}|)\frac{1}{2}\sum_{s=\pm}(1+s\sigma_{\bm{p}})n_F(E_{\bm{p}s}-\mu)\\
		=&-\int\frac{pdpd\theta}{(2\pi)^2}\frac{2\pi\alpha_{\delta}}{(p^2+k^2-2pk\cos\theta)^{\frac{1+\delta}{2}}}\left[\frac{1+\sigma_{\bm{k}}\cos(N\theta)}{2}n_F(\hbar v_fp-\mu)+\frac{1-\sigma_{\bm{k}}\cos(N\theta)}{2}n_F(-\hbar v_fp-\mu)\right]\\
\end{aligned}\end{equation}
where $\theta=\theta_{\bm{p}}-\theta_{\bm{k}}$. We have used the relation
\begin{equation}\begin{aligned}
		\sigma_{\bm{p}}=&\frac{p_+^N\sigma_++p_-^N\sigma_-}{p^N}=e^{iN\theta_{\bm{p}}}\sigma_++e^{-iN\theta_{\bm{p}}}\sigma_-=e^{iN\theta}e^{iN\theta_{\bm{k}}}\sigma_++e^{-iN\theta}e^{-iN\theta_{\bm{k}}}\sigma_-\\
		\to&(e^{iN\theta_{\bm{k}}}\sigma_++e^{-iN\theta_{\bm{k}}}\sigma_-)\cos(N\theta)=\sigma_{\bm{k}}\cos(N\theta)
\end{aligned}\end{equation}
where the terms containing $\sin N\theta$ can be omitted in the integral due to the parity analysis. Then we ignore the constant term and get
\begin{equation}\begin{aligned}\label{app:self-2}
		\Sigma(\bm{k},ik_n)=&\sigma_{\bm{k}}\frac{\alpha_{\delta}}{2\pi}\int_0^{\infty}pdp\int_{0}^{\pi}d\theta\frac{\cos(N\theta)}{(p^2+k^2-2pk\cos\theta)^{\frac{1+\delta}{2}}}
		=\sigma_{\bm{k}}\frac{\alpha_{\delta}}{2\pi}\frac{1}{\Gamma(\frac{1+\delta}{2})}\int_0^{\infty}dp\int_0^{\infty} dz\frac{p}{z^{(1-\delta)/2}}\int_{0}^{\pi}d\theta\cos(N\theta) e^{-(p^2+k^2-2pk\cos\theta)z}\\
		=&\sigma_{\bm{k}}\frac{\alpha_{\delta}}{2\pi}\frac{1}{\Gamma(\frac{1+\delta}{2})}\int_0^{\infty}dp\int_0^{\infty} dz\frac{p}{z^{(1-\delta)/2}}e^{-(p^2+k^2)z}\pi I_N(2kpz)
		=\sigma_{\bm{k}}\frac{\alpha_{\delta}}{2}\frac{1}{\Gamma(\frac{1+\delta}{2})}\int_0^{\infty}dz\frac{1}{z^{(1-\delta)/2}}\int_0^{\infty} dppe^{-(p^2+k^2)z} I_N(2kpz)\\
		=&\sigma_{\bm{k}}\frac{\alpha_{\delta}}{2}\frac{1}{\Gamma(\frac{1+\delta}{2})}\int_0^{\infty}dz\frac{1}{z^{(1-\delta)/2}}\frac{ke^{-\frac{k^2z}{2}}\sqrt{\pi}}{4\sqrt{z}}[I_{\frac{N-1}{2}}(\frac{k^2z}{2})+I_{\frac{N+1}{2}}(\frac{k^2z}{2})]\\
		=&k\sigma_{\bm{k}}\frac{\alpha_{\delta}}{8}\frac{\sqrt{\pi}}{\Gamma(\frac{1+\delta}{2})}\int_0^{\infty}dz\frac{e^{-\frac{k^2z}{2}}}{z^{(2-\delta)/2}}[I_{\frac{N-1}{2}}(\frac{k^2z}{2})+I_{\frac{N+1}{2}}(\frac{k^2z}{2})]
		=k\sigma_{\bm{k}}\frac{\alpha_{\delta}}{8}\frac{\sqrt{\pi}}{\Gamma(\frac{1+\delta}{2})}\frac{2^{\delta/2}}{k^{\delta}}\int_0^{\infty}dz'\frac{e^{-z'}}{z'^{(2-\delta)/2}}[I_{\frac{N-1}{2}}(z')+I_{\frac{N+1}{2}}(z')]\\
		=&k\sigma_{\bm{k}}\frac{\alpha_{\delta}}{8}\frac{\sqrt{\pi}}{\Gamma(\frac{1+\delta}{2})}\frac{2^{\delta/2}}{k^{\delta}}\frac{2^{-\delta/2}N\Gamma(\frac{1-\delta}{2})\Gamma(\frac{N-1+\delta}{2})}{\sqrt{\pi}\Gamma(\frac{N+3-\delta}{2})}
		=\frac{\alpha_{\delta}N}{8}\frac{\Gamma(\frac{1-\delta}{2})}{\Gamma(\frac{1+\delta}{2})}\frac{\Gamma(\frac{N-1+\delta}{2})}{\Gamma(\frac{N+3-\delta}{2})}k^{1-\delta}\sigma_{\bm{k}}
		=\phi(k)\sigma_{\bm{k}}
\end{aligned}\end{equation}
where $\alpha_{\delta}=\alpha_0r_0^{-\delta}2^{\delta}\frac{\Gamma(\frac{1+\delta}{2})}{\Gamma(\frac{1-\delta}{2})}$ and $\phi(k)=N\alpha_{0}r_0^{-\delta}2^{\delta-3}\frac{\Gamma(\frac{n-1+\delta}{2})}{\Gamma(\frac{n+3-\delta}{2})}k^{1-\delta}$. In the above derivation, we have used the identity:
\begin{equation}
	|k|^{-(1+\delta)}=\frac{1}{\Gamma(\frac{1+\delta}{2})}\int_0^{\infty}dz\frac{e^{-k^2z}}{z^{(1-\delta)/2}}
	\ \ \ \ \text{and}\ \ \ \ 
	I_N(z)=\frac{1}{\pi}\int_0^{\pi}e^{z\cos\theta}\cos(N\theta)d\theta
\end{equation}

Then, the self-energy correction for the correlation function is given by
\begin{equation}\begin{aligned}\label{eq:cor-self-2}
		C^{\text{self}}_{xyxy}(i\Omega)=&-8\int\frac{d^2\bm{k}}{(2\pi)^2}\frac{d\omega}{2\pi}{\rm Tr}[\mathcal{G}_{\bm{k}i(\omega+\Omega)}T_{xy}^{(0)}(\bm{k})\mathcal{G}_{\bm{k},i\omega}\Sigma(\bm{k})\mathcal{G}_{\bm{k},i\omega}T_{xy}^{(0)}(\bm{k})]\\
		=&-8\int\frac{d^2\bm{k}}{(2\pi)^2}\frac{d\omega}{2\pi}
		{\rm Tr}\left[\frac{1}{2}\sum_{s=\pm}\frac{1+s\sigma_{\bm{k}}}{i(\omega+\Omega)-s\alpha k^N}\right](i\frac{N\zeta_N}{2})(k_-k_+^{N-1}\sigma_+-k_+k_-^{N-1}\sigma_-)\left[\frac{1}{2}\sum_{s=\pm}\frac{1+s\sigma_{\bm{k}}}{i\omega-s\zeta_N k^N}\right]\phi(k)\sigma_{\bm{k}}\\
		&\left[\frac{1}{2}\sum_{s=\pm}\frac{1+s\sigma_{\bm{k}}}{i\omega-s\zeta_N k^N}\right](i\frac{N\zeta_N}{2})(k_-k_+^{N-1}\sigma_+-k_+k_-^{N-1}\sigma_-)\\
		=&2(N\zeta_N)^2\int\frac{d^2\bm{k}}{(2\pi)^2}\frac{d\omega}{2\pi}
		{\rm Tr}\left[-\frac{i(\omega+\Omega)+\zeta_N k^N\sigma_{\bm{k}}}{(\omega+\Omega)^2+(\zeta_N k^N)^2}\right](k_-k_+^{N-1}\sigma_+-k_+k_-^{N-1}\sigma_-)\left[-\frac{i\omega+\zeta_N k^N\sigma_{\bm{k}}}{\omega^2+(\zeta_N k^N)^2}\right]\phi(k)\sigma_{\bm{k}}\\
		&\left[-\frac{i\omega+\zeta_N k^N\sigma_{\bm{k}}}{\omega^2+(\zeta_N k^N)^2}\right](k_-k_+^{N-1}\sigma_+-k_+k_-^{N-1}\sigma_-)\\
		=&-2(N\zeta_N)^2\int\frac{d^2\bm{k}}{(2\pi)^2}\frac{d\omega}{2\pi}\frac{\phi(k)k^{2N}}{[(\omega+\Omega)^2+(\zeta_N k^N)^2][\omega^2+(\zeta_N k^N)^2]^2}\\
		&{\rm Tr}\left\{
		[i(\omega+\Omega)+\zeta_N k^N(e^{iN\theta_{\bm{k}}}\sigma_++e^{-iN\theta_{\bm{k}}}\sigma_-)][e^{i(N-2)\theta_{\bm{k}}}\sigma_+-e^{-i(N-2)\theta_{\bm{k}}}\sigma_-][i\omega+\zeta_N k^n(e^{iN\theta_{\bm{k}}}\sigma_++e^{-iN\theta_{\bm{k}}}\sigma_-)]\right.\\
		&\left.(e^{iN\theta_{\bm{k}}}\sigma_++e^{-iN\theta_{\bm{k}}}\sigma_-)[i\omega+\alpha k^n(e^{iN\theta_{\bm{k}}}\sigma_++e^{-iN\theta_{\bm{k}}}\sigma_-)][e^{i(N-2)\theta_{\bm{k}}}\sigma_+-e^{-i(N-2)\theta_{\bm{k}}}\sigma_-]\right\}\\
		=&-2(N\zeta_N)^2\int\frac{d^2\bm{k}}{(2\pi)^2}\frac{d\omega}{2\pi}\frac{\phi(k)k^{2N}}{[(\omega+\Omega)^2+(\zeta_N k^N)^2][\omega^2+(\zeta_N k^N)^2]^2}\\
		&{\rm Tr}\left\{[i(\omega+\Omega)(e^{i(N-2)\theta_{\bm{k}}}\sigma_+-e^{-i(N-2)\theta_{\bm{k}}}\sigma_-)+\zeta_N k^N(e^{-i2\theta_{\bm{k}}}\sigma_-\sigma_+-e^{i2\theta_{\bm{k}}}\sigma_+\sigma_-)]\right.\\
		&[i\omega(e^{iN\theta_{\bm{k}}}\sigma_++e^{-iN\theta_{\bm{k}}}\sigma_-)+\alpha k^N(\sigma_-\sigma_++\sigma_+\sigma_-)]\\
		&\left.[i\omega(e^{i(N-2)\theta_{\bm{k}}}\sigma_+-e^{-i(N-2)\theta_{\bm{k}}}\sigma_-)+\zeta_N k^N(e^{-i2\theta_{\bm{k}}}\sigma_-\sigma_+-e^{i2\theta_{\bm{k}}}\sigma_+\sigma_-)]\right\}\\
		=&-2(N\zeta_N)^2\int\frac{d^2\bm{k}}{(2\pi)^2}\frac{d\omega}{2\pi}\frac{\phi(k)k^{2n}\times4\omega(\omega+\Omega)\alpha k^N}{[(\omega+\Omega)^2+(\zeta_N k^N)^2][\omega^2+(\zeta_N k^N)^2]^2}\\
		=&-8(N\zeta_N)^2\int\frac{kdk}{2\pi}\phi(k)\zeta_N k^{3N}\int_{-\infty}^{\infty}\frac{d\omega}{2\pi}\frac{\omega(\omega+\Omega)}{[(\omega+\Omega)^2+(\zeta_N k^N)^2][\omega^2+(\zeta_N k^N)^2]^2}\\
		=&-8(N\zeta_N)^2\int\frac{kdk}{2\pi}N\alpha_{0}r_0^{-\delta}2^{\delta-3}\frac{\Gamma(\frac{N-1+\delta}{2})}{\Gamma(\frac{N+3-\delta}{2})}k^{1-\delta}\alpha k^{3N}\frac{1}{2\pi}\frac{\pi[4(\alpha k^N)^2-\Omega^2]}{2\alpha k^n[4(\alpha k^N)^2+\Omega^2]^2}\\
		=&-(N\zeta_N)^2\frac{N\alpha_{0}r_0^{-\delta}2^{\delta-3}}{\pi}\frac{\Gamma(\frac{N-1+\delta}{2})}{\Gamma(\frac{N+3-\delta}{2})}\int dk\frac{k^{2(N+1)-\delta}[4(\zeta_N k^N)^2+\Omega^2-2\Omega^2]}{[4(\zeta_N k^N)^2+\Omega^2]^2}\\
		=&-(N\zeta_N)^2\frac{N\alpha_{0}r_0^{-\delta}2^{\delta-3}}{\pi}\frac{\Gamma(\frac{N-1+\delta}{2})}{\Gamma(\frac{N+3-\delta}{2})}\int dk\left\{\frac{[4(\zeta_N k^N)^2+\Omega^2-\Omega^2]k^{2-\delta}}{4(\zeta_N k^N)^2+\Omega^2}-\frac{2\Omega^2[4(\alpha k^N)^2+\Omega^2-\Omega^2]k^{2-\delta}}{[4(\alpha k^N)^2+\Omega^2]^2}\right\}\frac{1}{4\zeta_N^2}\\
		=&-(N\zeta_N)^2\frac{N\alpha_{0}r_0^{-\delta}2^{\delta-3}}{\pi}\frac{\Gamma(\frac{N-1+\delta}{2})}{\Gamma(\frac{N+3-\delta}{2})}\int dk\left\{k^{2-\delta}-\frac{3\Omega^2k^{2-\delta}}{4(\zeta_N k^N)^2+\Omega^2}+\frac{2\Omega^4k^{2-\delta}}{[4(\zeta_N k^N)^2+\Omega^2]^2}\right\}\frac{1}{4\zeta_N^2}\\
\end{aligned}\end{equation}
where $\phi(k)=\alpha_{0}r_0^{-\delta}2^{\delta-3}\frac{\Gamma(\frac{N-1+\delta}{2})}{\Gamma(\frac{N+3-\delta}{2})}k^{1-\delta}$, $\alpha_0=\alpha\hbar v$, and ${\rm Tr}[\sigma_0]=2$. It is noticed that the first term in the brace contributes to the real part of the correlation function. Thus, we only consider the other two terms in the brace. Furthermore, in the case of $N=1$, the above equation goes back to the result of Ref~\cite{Link2018}, so the following derivation is evaluated for the condition $N\geq2$, 
\begin{equation}\begin{aligned}
		Q_1^{\text{self}}=&(N\zeta_N)^2\frac{N\alpha_{0}r_0^{-\delta}2^{\delta-3}}{\pi}\frac{\Gamma(\frac{N-1+\delta}{2})}{\Gamma(\frac{N+3-\delta}{2})}\int dk\frac{3\Omega^2k^{2-\delta}}{4(\zeta_N k^N)^2+\Omega^2}\frac{1}{4\zeta_N^2}
		=(N\zeta_N)^2\frac{N\alpha_{0}r_0^{-\delta}2^{\delta-3}}{\pi}\frac{\Gamma(\frac{N-1+\delta}{2})}{\Gamma(\frac{N+3-\delta}{2})}\int \frac{d(\zeta_N k^N)}{N\zeta_N k^{N-1}}\frac{3\Omega^2k^{2-\delta}}{4(\zeta_N k^N)^2+\Omega^2}\frac{1}{4\zeta_N^2}\\
		=&\frac{3N^2\zeta_N^{\frac{-3+\delta}{N}}\alpha_{0}r_0^{-\delta}}{2^{5-\delta}\pi}\frac{\Gamma(\frac{N-1+\delta}{2})}{\Gamma(\frac{N+3-\delta}{2})}\int dr\frac{\Omega^2r^{d-3}}{4r^2+\Omega^2}
		=-\frac{3N^2\zeta_N^{\frac{-3+\delta}{N}}\alpha_{0}r_0^{-\delta}}{2^{5-\delta}\pi}\frac{\Gamma(\frac{N-1+\delta}{2})}{\Gamma(\frac{N+3-\delta}{2})}\frac{2^{1-d}\pi\Omega^d\csc(\frac{d\pi}{2})}{\Omega^2}\\
		=&-\frac{3N^2\zeta_N^{\frac{-3+\delta}{N}}\alpha_{0}r_0^{-\delta}}{2^{d+4-\delta}}\frac{\Gamma(\frac{N-1+\delta}{2})}{\Gamma(\frac{N+3-\delta}{2})}\csc(\frac{d\pi}{2})\Omega^{\frac{3-\delta}{N}}\\
\end{aligned}\end{equation}
\begin{equation}\begin{aligned}
		Q_2^{\text{self}}=&-(N\zeta_N)^2\frac{N\alpha_{0}r_0^{-\delta}2^{\delta-3}}{\pi}\frac{\Gamma(\frac{N-1+\delta}{2})}{\Gamma(\frac{N+3-\delta}{2})}\int dk\frac{2\Omega^4k^{2-\delta}}{[4(\zeta_N k^N)^2+\Omega^2]^2}\frac{1}{4\zeta_N^2}
		=-\frac{N^2\zeta_N^{\frac{-3+\delta}{N}}\alpha_{0}r_0^{-\delta}}{2^{5-\delta}\pi}\frac{\Gamma(\frac{N-1+\delta}{2})}{\Gamma(\frac{N+3-\delta}{2})}\int dk\frac{2\Omega^4r^{d-3}}{[4r^2+\Omega^2]^2}\\
		=&-\frac{N^2\zeta_N^{\frac{-3+\delta}{N}}\alpha_{0}r_0^{-\delta}}{2^{5-\delta}\pi}\frac{\Gamma(\frac{N-1+\delta}{2})}{\Gamma(\frac{N+3-\delta}{2})}\frac{2^{1-d}(d-4)\pi\Omega^d\csc(\frac{d\pi}{2})}{\Omega^2}\\
		=&-\frac{N^2\zeta_N^{\frac{-3+\delta}{N}}\alpha_{0}r_0^{-\delta}(d-4)}{2^{d+4-\delta}}\frac{\Gamma(\frac{N-1+\delta}{2})}{\Gamma(\frac{N+3-\delta}{2})}\csc(\frac{d\pi}{2})\Omega^{\frac{3-\delta}{N}}\\
\end{aligned}\end{equation}
where $r=\alpha k^N$ and $d=\frac{2N+3-\delta}{N}$. By doing $\delta\to0$, the self-energy corrected correlation function is given by
\begin{equation}
	C_{xyxy}^{\text{self}}(i\Omega)=Q_1^{\text{self}}(\delta\to0)+Q_2^{\text{self}}(\delta\to0)=\alpha_0\frac{N(N+3)}{2^{6+\frac{3}{N}}}\frac{\Gamma(\frac{N-1}{2})}{\Gamma(\frac{N+3}{2})}\csc(\frac{3\pi}{2N})\zeta_N^{-\frac{3}{N}}\Omega^{\frac{3}{N}}
\end{equation}
\newpage
\section{The vertex diagram}\label{appendix:vertex}
The vertex correction for the correlation function is given by
\begin{equation}\begin{aligned}\label{eq:cor-vertex-2}
		C^{\text{ver}}_{xyxy}(i\Omega)=&4\int\frac{d\omega}{2\pi}\int\frac{d\omega'}{2\pi}\int\frac{d^2\bm{p}}{(2\pi)^2}\int\frac{d^2\bm{k}}{(2\pi)^2} V(|\bm{k}-\bm{p}|){\rm Tr}\left[\mathcal{G}_{\bm{p},i\omega}T_{xy}^{(0)}(\bm{p})\mathcal{G}_{\bm{p},i(\omega+\Omega)}\mathcal{G}_{\bm{k},i(\omega'+\Omega)}T_{xy}^{(0)}(\bm{k})\mathcal{G}_{\bm{k},i\omega'}\right]\\
		=&4\int\frac{d\omega}{2\pi}\int\frac{d\omega'}{2\pi}\int\frac{d^2\bm{p}}{(2\pi)^2}\int\frac{d^2\bm{k}}{(2\pi)^2}\frac{2\pi\alpha_{\delta}}{|\bm{k}-\bm{p}|^{1+\delta}}{\rm Tr}\left\{
		\left[-\frac{i\omega+\zeta_N p^N\sigma_{\bm{p}}}{\omega^2+(\zeta_N p^N)^2}\right]
		(i\frac{N\zeta_N}{2})(p_-p_+^{N-1}\sigma_+-p_+p_-^{N-1}\sigma_-)\right.\\
		&\left.\left[-\frac{i(\omega+\Omega)+\zeta_N p^N\sigma_{\bm{p}}}{(\omega+\Omega)^2+(\zeta_N p^N)^2}\right]
		\left[-\frac{i(\omega'+\Omega)+\zeta_N k^N\sigma_{\bm{k}}}{(\omega'+\Omega)^2+(\zeta_N k^N)^2}\right]
		(i\frac{N\zeta_N}{2})(k_-k_+^{N-1}\sigma_+-k_+k_-^{N-1}\sigma_-)
		\left[-\frac{i\omega'+\zeta_N k^N\sigma_{\bm{k}}}{\omega'^2+(\zeta_N k^N)^2}\right]\right\}\\
		=&-(N\zeta_N)^2\int\frac{d\omega}{2\pi}\int\frac{d\omega'}{2\pi}\int\frac{d^2\bm{p}}{(2\pi)^2}\int\frac{d^2\bm{k}}{(2\pi)^2}\frac{2\pi\alpha_{\delta}}{|\bm{k}-\bm{p}|^{1+\delta}}\\
		&\frac{1}{[\omega^2+(\zeta_N p^N)^2][(\omega+\Omega)^2+(\zeta_N p^N)^2][(\omega'+\Omega)^2+(\zeta_N k^N)^2][\omega'^2+(\zeta_N k^N)^2]}\\
		&{\rm Tr}\left\{[i\omega(p_-p_+^{N-1}\sigma_+-p_+p_-^{N-1}\sigma_-)+\zeta_N(p_-^{N+1}p_+^{N-1}\sigma_-\sigma_+-p_+^{N+1}p_-^{N-1}\sigma_+\sigma_-)]\right.\\
		&[-(\omega+\Omega)(\omega'+\Omega)+i(\omega+\Omega)\zeta_N(k_+^N\sigma_++k_-^N\sigma_-)+i(\omega'+\Omega)\zeta_N(p_+^N\sigma_++p_-^N\sigma_-)+\zeta_N^2(p_+^Nk_-^N\sigma_+\sigma_-+p_-^Nk_+^N\sigma_-\sigma_+)]\\
		&[i\omega'(k_-k_+^{N-1}\sigma_+-k_+k_-^{N-1}\sigma_-)+\zeta_N(k_-^{N+1}k_+^{N-1}\sigma_+\sigma_--k_+^{N+1}k_-^{N-1}\sigma_-\sigma_+)]\\
		=&(N\zeta_N)^2\int\frac{d\omega}{2\pi}\int\frac{d\omega'}{2\pi}\int\frac{d^2\bm{p}}{(2\pi)^2}\int\frac{d^2\bm{k}}{(2\pi)^2}\frac{2\pi\alpha_{\delta}}{|\bm{k}-\bm{p}|^{1+\delta}}\\
		&\frac{\omega\omega'(\omega+\Omega)(\omega'+\Omega)p^Nk^N[e^{i(N-2)\theta}+e^{-i(N-2)\theta}]-[\omega\omega'+(\omega+\Omega)(\omega'+\Omega)]\zeta_N^2p^{2N}k^{2N}(e^{2i\theta}+e^{-2i\theta})+\zeta_N^4p^{3N}k^{3N}[e^{i(N+2)\theta}+e^{-i(N+2)\theta}]}{[\omega^2+(\zeta_N p^N)^2][(\omega+\Omega)^2+(\zeta_N p^N)^2][(\omega'+\Omega)^2+(\zeta_N k^N)^2][\omega'^2+(\zeta_N k^N)^2]}\\
		=&(N\zeta_N)^2\int\frac{d^2\bm{p}}{(2\pi)^2}\int\frac{d^2\bm{k}}{(2\pi)^2}\frac{2\pi\alpha_{\delta}}{|\bm{k}-\bm{p}|^{1+\delta}}\\
		&\left\{\frac{2\zeta_N^2p^{2N}k^{2N}\cos[(N-2)\theta]}{[4(\zeta_N p^N)^2+\Omega^2][4(\zeta_N k^N)^2+\Omega^2]}-\frac{\Omega^2p^{n}k^{n}\cos(2\theta)}{[4(\zeta_N p^N)^2+\Omega^2][4(\zeta_N k^N)^2+\Omega^2]}+\frac{2\zeta_N^2p^{2N}k^{2N}\cos[(N+2)\theta]}{[4(\zeta_N p^N)^2+\Omega^2][4(\zeta_N k^N)^2+\Omega^2]}\right\}\\
		=&(N\zeta_N)^2\int\frac{d^2\bm{p}}{(2\pi)^2}\int\frac{d^2\bm{k}}{(2\pi)^2}\frac{2\pi\alpha_{\delta}}{|\bm{k}-\bm{p}|^{1+\delta}}
		\frac{4\zeta_N^2p^{2N}k^{2N}\cos(2\theta)\cos(N\theta)-\Omega^2p^{N}k^{N}\cos(2\theta)}{[4(\zeta_N p^N)^2+\Omega^2][4(\zeta_N k^N)^2+\Omega^2]}\\
		=&\frac{(N\zeta_N)^2\alpha_{\delta}}{2\pi^2}\int_0^{\infty} dp\int_0^{\infty} dk\int_0^{\pi}d\theta\frac{pk}{(p^2+k^2-2pk\cos\theta)^{\frac{1+\delta}{2}}}
		\frac{4\zeta_N^2p^{2N}k^{2N}\cos(2\theta)\cos(N\theta)-\Omega^2p^{N}k^{N}\cos(2\theta)}{[4(\zeta_N p^N)^2+\Omega^2][4(\zeta_N k^N)^2+\Omega^2]}\\
		=&\frac{(N\zeta_N)^2\alpha_{\delta}}{2\pi^2}\int_0^{\infty} dp\int_0^{\infty} dk\int_0^{\pi}d\theta\frac{pk}{(p^2+k^2-2pk\cos\theta)^{\frac{1+\delta}{2}}}\\
		&\left\{\frac{[4(\zeta_N p^{N})^2+\Omega^2][4(\zeta_N k^{N})^2+\Omega^2]-\Omega^2[4(\zeta_N p^{N})^2+\Omega^2]-\Omega^2[4(\zeta_N k^{N})^2+\Omega^2]+\Omega^4}{[4(\zeta_N p^N)^2+\Omega^2][4(\zeta_N k^N)^2+\Omega^2]}\frac{\cos(2\theta)\cos(N\theta)}{4\zeta_N^2}\right.\\
		&\left.-\frac{\Omega^2p^{N}k^{N}\cos(2\theta)}{[4(\zeta_N p^N)^2+\Omega^2][4(\zeta_N k^N)^2+\Omega^2]}\right\}\\
		=&\frac{(N\zeta_N)^2\alpha_{\delta}}{2\pi^2}\int_0^{\infty} dp\int_0^{\infty} dk\int_0^{\pi}d\theta\frac{pk}{(p^2+k^2-2pk\cos\theta)^{\frac{1+\delta}{2}}}\\
		&\left\{[1-\frac{2\Omega^2}{4(\zeta_N k^N)^2+\Omega^2}+\frac{\Omega^4}{[4(\zeta_N p^N)^2+\Omega^2][4(\zeta_N k^N)^2+\Omega^2]}]\frac{\cos(2\theta)\cos(N\theta)}{4\zeta_N^2}-\frac{\Omega^2p^{N}k^{N}\cos(2\theta)}{[4(\zeta_N p^N)^2+\Omega^2][4(\zeta_N k^N)^2+\Omega^2]}\right\}\\
\end{aligned}\end{equation}
We ignore the term in the brace that only contribute to the real part like the derivation of the self-energy correction so that the above equation can be calculated dividing into three parts: $Q^{\text{ver}}_1$, $Q_2^{\text{ver}}$, and $Q_3^{\text{ver}}$.
\newpage
\begin{equation}\begin{aligned}
		Q^{\text{ver}}_1=&-\frac{(N\zeta_N)^2\alpha_{\delta}}{2\pi^2}\int_0^{\infty} dp\int_0^{\infty} dk\int_0^{\pi}d\theta\frac{pk}{(p^2+k^2-2pk\cos\theta)^{\frac{1+\delta}{2}}}\frac{\Omega^2p^{N}k^{N}\cos(2\theta)}{[4(\zeta_N p^N)^2+\Omega^2][4(\zeta_N k^N)^2+\Omega^2]}\\
		=&-\frac{(N\zeta_N)^2\alpha_{\delta}}{2\pi^2}\int_0^{\infty} kdx\int_0^{\infty} dk\int_0^{\pi}d\theta\frac{xk^2}{k^{1+\delta}(x^2+1-2x\cos\theta)^{\frac{1+\delta}{2}}}\frac{\Omega^2x^{N}k^{2N}\cos(2\theta)}{[4(\zeta_N k^N)^2x^{2N}+\Omega^2][4(\zeta_N k^N)^2+\Omega^2]}\\
		=&-\frac{(N\zeta_N)^2\alpha_{\delta}}{2\pi^2}\int_0^{\infty} dx\int_0^{\infty} dk\int_0^{\pi}d\theta\frac{x^{N+1}\cos(2\theta)}{(x^2+1-2x\cos\theta)^{\frac{1+\delta}{2}}}\frac{\Omega^2k^{2N+2-\delta}}{[4(\zeta_N k^N)^2x^{2N}+\Omega^2][4(\zeta_N k^N)^2+\Omega^2]}\\
		=&-\frac{N\zeta_N^{\frac{-3+\delta}{N}}\alpha_{\delta}}{2\pi^2}\int_0^{\infty} dx\int_0^{\pi}d\theta\frac{x^{N+1}\cos(2\theta)}{(x^2+1-2x\cos\theta)^{\frac{1+\delta}{2}}}\int_0^{\infty} dr\frac{\Omega^2r^{d-1}}{[4r^2x^{2N}+\Omega^2][4r^2+\Omega^2]}\\
		=&-\frac{N\zeta_N^{\frac{-3+\delta}{N}}\alpha_{\delta}}{2\pi^2}\int_0^{\infty} dx\int_0^{\pi}d\theta\frac{x^{N+1}\cos(2\theta)}{(x^2+1-2x\cos\theta)^{\frac{1+\delta}{2}}}\frac{2^{-1-d}\pi[-\Omega^d+(x^{2N})^{1-d/2}\Omega^{d}]\csc(\frac{d\pi}{2})}{(x^{2N}-1)\Omega^2}\\
		=&\frac{N\zeta_N^{\frac{-3+\delta}{N}}\alpha_{\delta}}{2^{d+2}\pi}\csc(\frac{d\pi}{2})\Omega^{\frac{3-\delta}{N}}\left[\int_0^{\infty} dx\int_0^{\pi}d\theta\frac{x^{N+1}\cos(2\theta)}{(x^2+1-2x\cos\theta)^{\frac{1+\delta}{2}}}\frac{1-x^{-3+\delta}}{x^{2N}-1}\right]\\
		=&\frac{Nr_0^{-\delta}}{2^{d+2-\delta}\pi}\frac{\Gamma(\frac{1+\delta}{2})}{\Gamma(\frac{1-\delta}{2})}\csc(\frac{d\pi}{2})\left[\int_0^{\infty} dx\int_0^{\pi}d\theta\frac{x^{N+1}\cos(2\theta)}{(x^2+1-2x\cos\theta)^{\frac{1+\delta}{2}}}\frac{1-x^{-3+\delta}}{x^{2N}-1}\right]\zeta_N^{\frac{-3+\delta}{N}}\alpha_{0}\Omega^{\frac{3-\delta}{N}}\\
		\xrightarrow{\delta\to0}&-\alpha_{0}\frac{N}{2^{4+\frac{3}{N}}\pi}\csc(\frac{3\pi}{2N})f_1(N)\zeta_N^{-\frac{3}{N}}\Omega^{\frac{3}{N}}
\end{aligned}\end{equation}
with
\begin{equation}
	f_1(N)=\int_0^{\infty} dx\int_0^{\pi}d\theta\frac{x^{N+1}\cos(2\theta)}{(x^2+1-2x\cos\theta)^{\frac{1}{2}}}\frac{1-x^{-3}}{x^{2N}-1}
\end{equation}
where $\alpha_{\delta}=\alpha_0r_0^{-\delta}2^{\delta}\frac{\Gamma(\frac{1+\delta}{2})}{\Gamma(\frac{1-\delta}{2})}$, $x=\frac{p}{k}$, $r=\alpha k^N$ and $d=\frac{2N+3-\delta}{N}$. 

\begin{equation}\begin{aligned}
		Q_2^{\text{ver}}=&\frac{(N\zeta_N)^2\alpha_{\delta}}{2\pi^2}\int_0^{\infty} dp\int_0^{\infty} dk\int_0^{\pi}d\theta\frac{pk}{(p^2+k^2-2pk\cos\theta)^{\frac{1+\delta}{2}}}\frac{\Omega^4}{[4(\zeta_N p^N)^2+\Omega^2][4(\zeta_N k^N)^2+\Omega^2]}\frac{\cos(2\theta)\cos(N\theta)}{4\zeta_N^2}\\
		=&\frac{(N\zeta_N)^2\alpha_{\delta}}{2\pi^2}\int_0^{\infty} kdx\int_0^{\infty} dk\int_0^{\pi}d\theta\frac{xk^2}{k^{1+\delta}(x^2+1-2x\cos\theta)^{\frac{1+\delta}{2}}}\frac{\Omega^4}{[4(\zeta_N k^N)^2x^{2n}+\Omega^2][4(\zeta_N k^N)^2+\Omega^2]}\frac{\cos(2\theta)\cos(N\theta)}{4\zeta_N^2}\\
		=&\frac{(N\zeta_N)^2\alpha_{\delta}}{2\pi^2}\int_0^{\infty} dx\int_0^{\infty} dk\int_0^{\pi}d\theta\frac{x}{(x^2+1-2x\cos\theta)^{\frac{1+\delta}{2}}}\frac{\cos(2\theta)\cos(N\theta)}{4\zeta_N^2}\frac{\Omega^4k^{2-\delta}}{[4(\zeta_N k^N)^2x^{2N}+\Omega^2][4(\zeta_N k^N)^2+\Omega^2]}\\
		=&\frac{N\zeta_N^{\frac{-3+\delta}{N}}\alpha_{\delta}}{8\pi^2}\int_0^{\infty} dx\int_0^{\pi}d\theta\frac{x\cos(2\theta)\cos(N\theta)}{(x^2+1-2x\cos\theta)^{\frac{1+\delta}{2}}}\int_0^{\infty} dr\frac{\Omega^4r^{d-3}}{[4r^2x^{2N}+\Omega^2][4r^2+\Omega^2]}\\
		=&\frac{N\zeta_N^{\frac{-3+\delta}{N}}\alpha_{\delta}}{8\pi^2}\int_0^{\infty} dx\int_0^{\pi}d\theta\frac{x\cos(2\theta)\cos(N\theta)}{(x^2+1-2x\cos\theta)^{\frac{1+\delta}{2}}}\frac{2^{1-d}\pi[\Omega^d-(x^{2N})^{2-d/2}\Omega^{d}]\csc(\frac{d\pi}{2})}{(x^{2N}-1)\Omega^2}\\
		=&\frac{N\zeta_N^{\frac{-3+\delta}{N}}\alpha_{\delta}}{2^{d+2}\pi}\csc(\frac{d\pi}{2})\Omega^{\frac{3-\delta}{N}}\left[\int_0^{\infty} dx\int_0^{\pi}d\theta\frac{x\cos(2\theta)\cos(N\theta)}{(x^2+1-2x\cos\theta)^{\frac{1+\delta}{2}}}\frac{1-x^{2N-3+\delta}}{x^{2N}-1}\right]\\
		=&\frac{Nr_0^{-\delta}}{2^{d+2-\delta}\pi}\frac{\Gamma(\frac{1+\delta}{2})}{\Gamma(\frac{1-\delta}{2})}\csc(\frac{d\pi}{2})\left[\int_0^{\infty} dx\int_0^{\pi}d\theta\frac{x\cos(2\theta)\cos(N\theta)}{(x^2+1-2x\cos\theta)^{\frac{1+\delta}{2}}}\frac{1-x^{2N-3+\delta}}{x^{2N}-1}\right]\zeta_N^{\frac{-3+\delta}{N}}\alpha_{0}\Omega^{\frac{3-\delta}{N}}\\
		\xrightarrow{\delta\to0}&-\alpha_{0}\frac{N}{2^{4+\frac{3}{N}}\pi}\csc(\frac{3\pi}{2N})f_2(N)\zeta_N^{-\frac{3}{N}}\Omega^{\frac{3}{N}}
\end{aligned}\end{equation}
with
\begin{equation}
	f_2(N)=\int_0^{\infty} dx\int_0^{\pi}d\theta\frac{x\cos(2\theta)\cos(N\theta)}{(x^2+1-2x\cos\theta)^{\frac{1}{2}}}\frac{1-x^{2N-3}}{x^{2N}-1}
\end{equation}
\begin{equation}\begin{aligned}
		Q_3^{\text{ver}}=&-\frac{(N\zeta_N)^2\alpha_{\delta}}{2\pi^2}\int_0^{\infty} dp\int_0^{\infty} dk\int_0^{\pi}d\theta\frac{pk}{(p^2+k^2-2pk\cos\theta)^{\frac{1+\delta}{2}}}\frac{2\Omega^2}{4(\zeta_N k^N)^2+\Omega^2}\frac{\cos(2\theta)\cos(N\theta)}{4\zeta_N^2}\\
		=&-\frac{(N\zeta_N)^2\alpha_{\delta}}{2\pi^2}\int_0^{\infty} dx\int_0^{\infty} dk\int_0^{\pi}d\theta\frac{x}{(x^2+1-2x\cos\theta)^{\frac{1+\delta}{2}}}\frac{\cos(2\theta)\cos(N\theta)}{4\zeta_N^2}\frac{2\Omega^2k^{2-\delta}}{4(\zeta_N k^N)^2+\Omega^2}\\
		=&-\frac{N\zeta_N^{\frac{-3+\delta}{N}}\alpha_{\delta}}{4\pi^2}\int_0^{\infty} dx\int_0^{\pi}d\theta\frac{x\cos(2\theta)\cos(N\theta)}{(x^2+1-2x\cos\theta)^{\frac{1+\delta}{2}}}\int_0^{\infty}dr\frac{\Omega^2r^{d-3}}{4r^2+\Omega^2}\\
		=&\frac{N\zeta_N^{\frac{-3+\delta}{N}}\alpha_{\delta}}{4\pi^2}\int_0^{\infty} dx\int_0^{\pi}d\theta\frac{x\cos(2\theta)\cos(N\theta)}{(x^2+1-2x\cos\theta)^{\frac{1+\delta}{2}}}\frac{2^{1-d}\pi\Omega^{d}\csc(\frac{d\pi}{2})}{\Omega^2}\\
		=&\frac{N\zeta_N^{\frac{-3+\delta}{N}}\alpha_{\delta}}{2^{d+1}\pi}\csc(\frac{d\pi}{2})\Omega^{\frac{3-\delta}{N}}\left[\int_0^{\infty} dx\int_0^{\pi}d\theta\frac{x\cos(2\theta)\cos(N\theta)}{(x^2+1-2x\cos\theta)^{\frac{1+\delta}{2}}}\right]\\
		=&\frac{N\zeta_N^{\frac{-3+\delta}{N}}\alpha_{\delta}}{2^{d+1}\pi}\csc(\frac{d\pi}{2})\Omega^{\frac{3-\delta}{N}}\int_0^{\infty} xdx\int_0^{\pi}d\theta \frac{\cos[(N+2)\theta]+\cos[(N-2)\theta]}{2}\frac{1}{\Gamma(\frac{1+\delta}{2})}\int_0^{\infty}dz\frac{e^{-(x^2+1-2x\cos\theta)z}}{z^{(1-\delta)/2}}\\
		=&\frac{N\zeta_N^{\frac{-3+\delta}{N}}\alpha_{\delta}}{2^{d+2}}\csc(\frac{d\pi}{2})\Omega^{\frac{3-\delta}{N}}\frac{1}{\Gamma(\frac{1+\delta}{2})}\int_0^{\infty}dz\frac{e^{-z}}{z^{(1-\delta)/2}}\int_0^{\infty} dxxe^{-x^2z}[I_{N+2}(2xz)+I_{N-2}(2xz)]\\
		=&\frac{N\zeta_N^{\frac{-3+\delta}{N}}\alpha_{\delta}}{2^{d+2}}\csc(\frac{d\pi}{2})\Omega^{\frac{3-\delta}{N}}\frac{1}{\Gamma(\frac{1+\delta}{2})}\int_0^{\infty}dz\frac{e^{-z}}{z^{(1-\delta)/2}}\frac{e^{z/2}\sqrt{\pi}}{2z^{3/2}}[(z-1+N)I_{\frac{N-1}{2}}(\frac{z}{2})+(z-1-N)I_{\frac{N+1}{2}}(\frac{z}{2})]\\
		=&\frac{N\zeta_N^{\frac{-3+\delta}{N}}\alpha_{\delta}\sqrt{\pi}}{2^{d+3}}\csc(\frac{d\pi}{2})\Omega^{\frac{3-\delta}{N}}\frac{1}{\Gamma(\frac{1+\delta}{2})}\int_0^{\infty}dz\frac{e^{-z/2}}{z^{2-\delta/2}}[(z-1+N)I_{\frac{N-1}{2}}(\frac{z}{2})+(z-1-N)I_{\frac{N+1}{2}}(\frac{z}{2})]\\
		=&\frac{Nr_0^{-\delta}\sqrt{\pi}}{2^{d+3-\delta}}\csc(\frac{d\pi}{2})\frac{1}{\Gamma(\frac{1-\delta}{2})}\left\{\int_0^{\infty}dz\frac{e^{-z/2}}{z^{2-\delta/2}}[(z-1+N)I_{\frac{N-1}{2}}(\frac{z}{2})+(z-1-N)I_{\frac{N+1}{2}}(\frac{z}{2})]\right\}\zeta_N^{\frac{-3+\delta}{N}}\alpha_0\Omega^{\frac{3-\delta}{N}}\\
		=&\alpha_{0}\frac{Nr_0^{-\delta}\sqrt{\pi}}{2^{d+3-\delta}}\frac{1}{\Gamma(\frac{1-\delta}{2})}\csc(\frac{d\pi}{2})g_1(N)\zeta_N^{\frac{-3+\delta}{N}}\Omega^{\frac{3-\delta}{N}}\\
		\xrightarrow{\delta\to0}&-\alpha_{0}\frac{N}{2^{5+\frac{3}{N}}}\csc(\frac{3\pi}{2N})g_1(N)\zeta_N^{-\frac{3}{N}}\Omega^{\frac{3}{N}}
\end{aligned}\end{equation}
with
\begin{equation}
	g_1(N)=\int_0^{\infty}dz\frac{e^{-z/2}}{z^{2}}[(z-1+N)I_{\frac{N-1}{2}}(\frac{z}{2})+(z-1-N)I_{\frac{N+1}{2}}(\frac{z}{2})]
\end{equation}

\newpage
\section{The honey diagram}\label{appendix:honey-correction}
\begin{equation}\begin{aligned}
		C_{xyxy}^{\text{honey}}(i\Omega)=&16\int\frac{d^2\bm{k}}{(2\pi)^2}\int\frac{d^2\bm{l}}{(2\pi)^2}\int\frac{d\omega'}{2\pi}\int\frac{d\omega}{2\pi}T_{xy}^{\text{int}}(\bm{l}){\rm Tr}\left[\mathcal{G}_{\bm{k}+\bm{l},i\omega'}\mathcal{G}_{\bm{k},i(\omega+\Omega)}T_{xy}^{(0)}(\bm{k})\mathcal{G}_{\bm{k},i\omega}\right]\\
		=&-(1-\delta)\alpha_0r_0^{-\delta}2^{\delta+4}\pi\frac{\Gamma(\frac{3+\delta}{2})}{\Gamma(\frac{3-\delta}{2})}\int\frac{d^2\bm{k}}{(2\pi)^2}\int\frac{d^2\bm{l}}{(2\pi)^2}\int\frac{d\omega'}{2\pi}\int\frac{d\omega}{2\pi}\frac{l_xl_y}{l^{3+\delta}}{\rm Tr}\\
		&\left[-\frac{i\omega'+\zeta_N(p_+^N\sigma_++p_-^N\sigma_-)}{\omega'^2+(\zeta_N p^N)^2}\right]
		\left[-\frac{i(\omega+\Omega)+\zeta_N(k_+^N\sigma_++k_-^N\sigma_-)}{(\omega+\Omega)^2+(\zeta_N k^N)^2}\right]
		(i\frac{N\zeta_N}{2})[k_-k_+^{N-1}\sigma_+-k_+k_-^{N-1}\sigma_-]
		\left[-\frac{i\omega+\zeta_N(k_+^N\sigma_++k_-^N\sigma_-)}{\omega^2+(\zeta_N k^N)^2}\right]\\
		=&(1-\delta)\alpha_0r_0^{-\delta}2^{\delta+4}\pi\frac{\Gamma(\frac{3+\delta}{2})}{\Gamma(\frac{3-\delta}{2})}\int\frac{d^2\bm{k}}{(2\pi)^2}\int\frac{d^2\bm{l}}{(2\pi)^2}\int\frac{d\omega'}{2\pi}\int\frac{d\omega}{2\pi}(i\frac{N\zeta_N}{2})\frac{l_xl_y}{l^{3+\delta}}\\
		&\frac{\omega(\omega+\Omega)\zeta_N(p_+^Nk_+k_-^{N-1}-p_-^Nk_-k_+^{N-1})+\zeta_N^3(p_+^Nk_-^{2N+1}k_+^{N-1}-p_-^Nk_+^{2N+1}k_-^{N-1})}{[\omega'^2+(\zeta_N p^N)^2][(\omega+\Omega)^2+(\zeta_N k^N)^2][\omega^2+(\zeta_N k^N)^2]}\\
		=&(1-\delta)\alpha_0r_0^{-\delta}2^{\delta+3}\pi\frac{\Gamma(\frac{3+\delta}{2})}{\Gamma(\frac{3-\delta}{2})}\int\frac{d^2\bm{k}}{(2\pi)^2}\int\frac{d^2\bm{l}}{(2\pi)^2}\int\frac{d\omega}{2\pi}(i\frac{N\zeta_N}{2})\frac{l_xl_y}{l^{3+\delta}}\\
		&\frac{\omega(\omega+\Omega)(p_+^Nk_+k_-^{N-1}-p_-^Nk_-k_+^{N-1})+\zeta_N^2(p_+^Nk_-^{2N+1}k_+^{N-1}-p_-^Nk_+^{2N+1}k_-^{N-1})}{p^N[(\omega+\Omega)^2+(\zeta_N k^N)^2][\omega^2+(\zeta_N k^N)^2]}\\
		=&(1-\delta)\alpha_0r_0^{-\delta}2^{\delta+3}\pi\frac{\Gamma(\frac{3+\delta}{2})}{\Gamma(\frac{3-\delta}{2})}\int\frac{d^2\bm{k}}{(2\pi)^2}\int\frac{d^2\bm{l}}{(2\pi)^2}(i\frac{N\zeta_N}{2})\frac{l_xl_y}{l^{3+\delta}}\\
		&\frac{\zeta_Nk^N(p_+^Nk_+k_-^{N-1}-p_-^Nk_-k_+^{N-1})+\zeta_N^2(\zeta_Nk^N)^{-1}(p_+^Nk_-^{2N+1}k_+^{N-1}-p_-^Nk_+^{2N+1}k_-^{N-1})}{p^N[4(\zeta_Nk^N)^2+\Omega^2]}\\
		=&(1-\delta)\alpha_0r_0^{-\delta}2^{\delta+3}\pi\frac{\Gamma(\frac{3+\delta}{2})}{\Gamma(\frac{3-\delta}{2})}\int\frac{d^2\bm{k}}{(2\pi)^2}\int\frac{d^2\bm{p}}{(2\pi)^2}(i\frac{N\zeta_N}{2})\frac{1}{4i}\frac{(p_+-k_+)^2-(p_--k_-)^2}{|\bm{p}-\bm{k}|^{3+\delta}}\\
		&\zeta_N\frac{k^{N+2}(p_+^Nk_-^{N-2}-p_-^Nk_+^{N-2})+k^{N-2}(p_+^Nk_-^{N+2}-p_-^Nk_+^{N+2})}{p^N[4(\zeta_N k^N)^2+\Omega^2]}\\
		=&(N\zeta_N^2)(1-\delta)\alpha_0r_0^{-\delta}2^{\delta}\pi\frac{\Gamma(\frac{3+\delta}{2})}{\Gamma(\frac{3-\delta}{2})}\int\frac{d^2\bm{k}}{(2\pi)^2}\int\frac{d^2\bm{p}}{(2\pi)^2}\frac{(p_+^2-p_-^2)+(k_+^2-k_-^2)-2p_+k_++2p_-k_-}{|\bm{p}-\bm{k}|^{3+\delta}}\\
		&\frac{k^{N+2}(p_+^Nk_-^{N-2}-p_-^Nk_+^{N-2})+k^{N-2}(p_+^Nk_-^{N+2}-p_-^Nk_+^{N+2})}{p^N[4(\zeta_N k^N)^2+\Omega^2]}\\
		=&(N\zeta_N^2)(1-\delta)\alpha_0r_0^{-\delta}2^{\delta}\pi\frac{\Gamma(\frac{3+\delta}{2})}{\Gamma(\frac{3-\delta}{2})}\int\frac{d^2\bm{k}}{(2\pi)^2}\int\frac{d^2\bm{p}}{(2\pi)^2}\\
		&\frac{p^2[e^{i(N+2)\theta}+e^{-i(N+2)\theta}-e^{i(N-2)\theta}-e^{-i(N-2)\theta}]-2pk[e^{i(N+1)\theta}+e^{-i(N+1)\theta}-e^{i(N-1)\theta}-e^{-i(N-1)\theta}]}{(p^2+k^2-2kp\cos\theta)^{\frac{3+\delta}{2}}}\frac{k^{2N}}{4(\zeta_N k^N)^2+\Omega^2}\\
		=&\frac{N\zeta_N^2}{2\pi^2}(1-\delta)\alpha_0r_0^{-\delta}2^{\delta}\frac{\Gamma(\frac{3+\delta}{2})}{\Gamma(\frac{3-\delta}{2})}\int kdk \int pdp\int_0^{\pi}d\theta\\
		&\frac{p^2[\cos(N+2)\theta-\cos(N-2)\theta]-2pk[\cos(N+1)\theta-\cos(N-1)\theta]}{(p^2+k^2-2kp\cos\theta)^{\frac{3+\delta}{2}}}\frac{k^{2N}}{4(\zeta_N k^N)^2+\Omega^2}\\
		=&\frac{N\zeta_N^2}{2\pi^2}(1-\delta)\alpha_0r_0^{-\delta}2^{\delta}\frac{\Gamma(\frac{3+\delta}{2})}{\Gamma(\frac{3-\delta}{2})}\int \frac{k^{2N+2-\delta}}{4(\zeta_N k^N)^2+\Omega^2}dk \int dx\int_0^{\pi}d\theta\\
		&\frac{x^3[\cos(N+2)\theta-\cos(N-2)\theta]-2x^2[\cos(N+1)\theta-\cos(N-1)\theta]}{(x^2+1-2x\cos\theta)^{\frac{3+\delta}{2}}}\\
		=&\frac{N\zeta_N^2}{2\pi^2}(1-\delta)\alpha_0r_0^{-\delta}2^{\delta}\frac{\Gamma(\frac{3+\delta}{2})}{\Gamma(\frac{3-\delta}{2})}\int dx\int_0^{\pi}d\theta
		\frac{x^3[\cos(N+2)\theta-\cos(N-2)\theta]-2x^2[\cos(N+1)\theta-\cos(N-1)\theta]}{(x^2+1-2x\cos\theta)^{\frac{3+\delta}{2}}}\\
		&\int dk \left[1-\frac{\Omega^2}{4(\zeta_N k^N)^2+\Omega^2}\right]\frac{k^{2-\delta}}{4\alpha^2}\\
\end{aligned}\end{equation}
where $\bm{p}=\bm{k}+\bm{l}$, $\theta=\theta_{\bm{p}}-\theta_{\bm{k}}$ and $x=\frac{p}{k}$. Consider the condition $N\geq2$ and ignoring the term only contribute to the real part, the above equation is 
\begin{equation}\begin{aligned}
		C_{xyxy}^{\text{honey}}(i\Omega)=&-\frac{N\zeta_N^2}{2\pi^2}(1-\delta)\alpha_0r_0^{-\delta}2^{\delta}\frac{\Gamma(\frac{3+\delta}{2})}{\Gamma(\frac{3-\delta}{2})}\int dx\int_0^{\pi}d\theta
		\frac{x^3[\cos(N+2)\theta-\cos(N-2)\theta]-2x^2[\cos(N+1)\theta-\cos(N-1)\theta]}{(x^2+1-2x\cos\theta)^{\frac{3+\delta}{2}}}\\
		&\int dk \frac{\Omega^2}{4(\zeta_N k^N)^2+\Omega^2}\frac{k^{2-\delta}}{4\zeta_N^2}\\
		=&-\frac{\zeta_N^{\frac{-3+\delta}{N}}}{8\pi^2}(1-\delta)\alpha_0r_0^{-\delta}2^{\delta}\frac{\Gamma(\frac{3+\delta}{2})}{\Gamma(\frac{3-\delta}{2})}\int dx\int_0^{\pi}d\theta
		\frac{x^3[\cos(N+2)\theta-\cos(N-2)\theta]-2x^2[\cos(N+2)\theta-\cos(N-2)\theta]}{(x^2+1-2x\cos\theta)^{\frac{3+\delta}{2}}}\int dr \frac{\Omega^2r^{d-3}}{4r^2+\Omega^2}\\
		=&\frac{\zeta_N^{\frac{-3+\delta}{N}}}{8\pi^2}(1-\delta)\alpha_0r_0^{-\delta}2^{\delta}\frac{\Gamma(\frac{3+\delta}{2})}{\Gamma(\frac{3-\delta}{2})}\int dx\int_0^{\pi}d\theta
		\frac{x^3[\cos(N+2)\theta-\cos(N-2)\theta]-2x^2[\cos(N+1)\theta-\cos(N-1)\theta]}{(x^2+1-2x\cos\theta)^{\frac{3+\delta}{2}}}\frac{2^{1-d}\pi\Omega^d\csc(\frac{d\pi}{2})}{\Omega^2}\\
		=&\frac{\zeta_N^{\frac{-3+\delta}{N}}}{2^{d+2}\pi}(1-\delta)\alpha_0r_0^{-\delta}2^{\delta}\frac{\Gamma(\frac{3+\delta}{2})}{\Gamma(\frac{3-\delta}{2})}\csc(\frac{d\pi}{2})\Omega^{\frac{3-\delta}{N}}\int dx\int_0^{\pi}d\theta
		\frac{x^3[\cos(N+2)\theta-\cos(N-2)\theta]-2x^2[\cos(N+1)\theta-\cos(N-1)\theta]}{(x^2+1-2x\cos\theta)^{\frac{3+\delta}{2}}}\\
		=&\frac{\zeta_N^{\frac{-3+\delta}{N}}}{2^{d+2}\pi}(1-\delta)\alpha_0r_0^{-\delta}2^{\delta}\frac{\Gamma(\frac{3+\delta}{2})}{\Gamma(\frac{3-\delta}{2})}\csc(\frac{d\pi}{2})\Omega^{\frac{3-\delta}{N}}\int dx\int_0^{\pi}d\theta
		\{x^3[\cos(N+2)\theta-\cos(N-2)\theta]-2x^2[\cos(N+1)\theta-\cos(N-1)\theta]\}\\
		&\frac{1}{\Gamma(\frac{3+\delta}{2})}\int_0^{\infty}dz\frac{e^{-(x^2+1-2x\cos\theta)z}}{z^{-(1+\delta)/2}}\\
		=&\frac{\zeta_N^{\frac{-3+\delta}{n}}}{2^{d+1}\pi}\alpha_0r_0^{-\delta}2^{\delta}\frac{1}{\Gamma(\frac{1-\delta}{2})}\csc(\frac{d\pi}{2})\Omega^{\frac{3-\delta}{N}}\int dx\int_0^{\pi}d\theta
		\{x^3[\cos(N+2)\theta-\cos(N-2)\theta]-2x^2[\cos(N+1)\theta-\cos(N-1)\theta]\}\\
		&\int_0^{\infty}dz\frac{e^{-(x^2+1-2x\cos\theta)z}}{z^{-(1+\delta)/2}}\\
		=&\frac{\alpha^{\frac{-3+\delta}{N}}}{2^{d+1}}\alpha_0r_0^{-\delta}2^{\delta}\frac{1}{\Gamma(\frac{1-\delta}{2})}\csc(\frac{d\pi}{2})\Omega^{\frac{3-\delta}{N}}\int_0^{\infty}dz\frac{e^{-z}}{z^{-(1+\delta)/2}}\int dx
		e^{-x^2z}\{x^3[I_{N+2}(2xz)-I_{N-2}(2xz)]-2x^2[I_{N+1}(2xz)-I_{N-1}(2xz)]\}\\
		=&\frac{\zeta_N^{\frac{-3+\delta}{N}}}{2^{d+1}}\alpha_0r_0^{-\delta}2^{\delta}\frac{1}{\Gamma(\frac{1-\delta}{2})}\csc(\frac{d\pi}{2})\Omega^{\frac{3-\delta}{N}}\int_0^{\infty}dz\frac{e^{-z}}{z^{-(1+\delta)/2}}\frac{Ne^{z/2}\sqrt{\pi}}{4z^{5/2}}[(1-N)I_{\frac{N-1}{2}}(\frac{z}{2})+(1+N)I_{\frac{N+1}{2}}(\frac{z}{2})]\\
		=&\alpha_0\frac{Nr_0^{-\delta}\sqrt{\pi}}{2^{d+3-\delta}}\frac{1}{\Gamma(\frac{1-\delta}{2})}\csc(\frac{d\pi}{2})g_2(N)\zeta_N^{\frac{-3+\delta}{N}}\Omega^{\frac{3-\delta}{N}}\\
		\xrightarrow{\delta\to0}&-\alpha_0\frac{N}{2^{5+\frac{3}{N}}}\csc(\frac{3\pi}{2N})g_2(N)\zeta_N^{-\frac{3}{N}}\Omega^{\frac{3}{N}}\\
\end{aligned}\end{equation}
with
\begin{equation}
	g_2(N)=\int_0^{\infty}dz\frac{e^{-z/2}}{z^{2}}[(1-N)I_{\frac{N-1}{2}}(\frac{z}{2})+(1+N)I_{\frac{N+1}{2}}(\frac{z}{2})]
\end{equation}
where $r=\alpha k^N$ and $d=\frac{2N+3-\delta}{N}$.

\end{appendix}
\end{widetext}


\end{document}